\documentclass[12pt]{article}

\usepackage{amsmath}
\usepackage{url}
\usepackage{epstopdf,gensymb}
\usepackage{graphics}
\usepackage{graphicx,color}
\usepackage{natbib}
\usepackage{amsfonts, animate,subfigure}
\usepackage{amssymb,dsfont}
\usepackage[margin=1in]{geometry}
\usepackage{setspace}
\usepackage[ruled]{algorithm2e}
\usepackage{bm}
\usepackage{subfigure}
\usepackage{authblk}

\usepackage[format=hang,labelfont=bf]{caption}

\usepackage{url}
\usepackage{hyperref}
\usepackage{todonotes}
\usepackage[capposition=top]{floatrow}
\usepackage{paralist}

\newcommand{\mbf}{\mathbf}

\usepackage[rightbars, color]{changebar}
\cbcolor{white}
\newcommand{\rev}[1]{\textcolor{black}{ #1}}
\newcommand{\revii}[1]{\cbstart{#1\cbend}}

\begin{document}

\title{Probabilistic Cause-of-death Assignment using Verbal Autopsies\thanks{Preparation of this manuscript was supported by the Bill and Melinda Gates Foundation, with partial support from a seed grant from the Center for the Studies of Demography and Ecology at the University of Washington along with grant K01 HD057246 to Clark and K01 HD078452 to McCormick, both from the National Institute of Child Health and Human Development (NICHD).  The authors are grateful to Peter Byass, Basia Zaba, Laina Mercer, Stephen Tollman, Adrian Raftery, Philip Setel, Osman Sankoh, and Jon Wakefield for helpful discussions.  We are also grateful to the MRC/Wits Rural Public Health and Health Transitions Research Unit and the Karonga Prevention Study for sharing their data for this project.}}

\author[1,2,3,*]{Tyler H. McCormick}
\author[1]{Zehang Richard Li}
\author[8,6]{Clara Calvert}
\author[6,8,9]{Amelia C. Crampin}
\author[5,7]{Kathleen Kahn}
\author[3,4,5,6,7]{Samuel J. Clark}

\affil[1]{Department of Statistics, University of Washington}
\affil[2]{Center for Statistics and the Social Sciences (CSSS), University of Washington}
\affil[3]{Department  of  Sociology, University of Washington}
\affil[4]{Institute of Behavioral  Science (IBS), University of Colorado at Boulder}
\affil[5]{MRC/Wits Rural Public Health and Health Transitions Research Unit (Agincourt), School of Public Health, Faculty of Health Sciences, University of the Witwatersrand}
\affil[6]{ALPHA Network, London}
\affil[7]{INDEPTH Network, Ghana}
\affil[8]{London School of Hygiene and Tropical Medicine}
\affil[9]{Karonga Prevention Study, Malawi}
\affil[*]{Correspondence to: \texttt{tylermc@uw.edu}}

\date{}
\maketitle

\begin{abstract}
\revii{In regions without complete-coverage} civil registration and vital statistics systems there is uncertainty about even the most basic demographic indicators.  In such regions the majority of deaths occur outside hospitals and are not recorded.  Worldwide, fewer than one-third of deaths are assigned a cause, with the least information available from the most impoverished nations.  In populations like this, verbal autopsy (VA) is a commonly used tool to assess cause of death and estimate cause-specific mortality rates and the distribution of deaths by cause.  VA uses an interview with caregivers of the decedent to elicit data describing the signs and symptoms leading up to the death.  This paper develops a new statistical tool known as \textit{InSilicoVA} to classify cause of death using information acquired through VA.  InSilicoVA shares uncertainty between cause of death assignments for specific individuals and the distribution of deaths by cause across the population.  Using side-by-side comparisons with both observed and simulated data, we demonstrate that InSilicoVA has distinct advantages compared to currently available methods. 
\end{abstract}



\doublespacing

\section{Introduction}
Data describing cause of death are critical to formulate, implement and evaluate public health policy.  Fewer than one-third of deaths worldwide are assigned a cause, with the most impoverished nations having the least information \citep{Horton2007}.  In 2007 {\em The Lancet} published a special issue titled ``Who Counts?''~\citep{AbouZahr2007,Boerma,Hill2007b,Horton2007,Mahapatra2007,Setel2007}; the authors identify the ``scandal of invisibility'' resulting from the lack of accurate, timely, full-coverage civil registration and vital statistics systems in much of the developing world.  They argue for a transformation in how civil registration is conducted in those parts of the world so that we are able to monitor health and design and evaluate effective interventions in a timely way.  \citet{Horton2007} argues that the past four decades have seen ``little progress''  and ``limited attention to vital registration'' by national and international organizations.  With bleak prospects for widespread civil registration in the coming decades, \citet{AbouZahr2007} recommends ``censuses and survey-based
approaches will have to be used to obtain the population representative data.''  This paper develops a statistical method for analyzing data based on one such survey.  The proposed method infers a likely cause of death for a given individual, while simultaneously estimating a population distribution of deaths by cause.  Individual deaths by cause can be related to surviving family members, while the population distribution of deaths provides critical information about the leading risks to population health.  Critically, the proposed method also provides a statistical framework for quantifying uncertainty in such data.

\subsection{Verbal autopsy}
We propose a method for using survey-based data to infer an individual's cause of death and a distribution of deaths by cause for the population.  The data are derived from verbal autopsy (VA) interviews. VA is conducted by administering a standardized questionnaire to caregivers, family members and/or others knowledgeable of the circumstances of a recent death.  The resulting data describe the decedent's health history leading up to death with a mixture of binary, numeric, categorical and narrative data.  These data describe the sequence and duration of important signs and symptoms leading up to the death.  The goal is to infer the likely causes of death from these data \citep{Byass:2012}.  VA has been widely used by researchers in Health and Demographic Surveillance System (HDSS) sites, such as members of the INDEPTH Network \citep{Sankoh2012} and the ALPHA Network~\citep{maher2010translating}, and has recently received renewed attention from the World Health Organization (WHO) through the release of an update to the widely-used WHO standard VA questionnaire~\citep{who}.  The main statistical challenge with VA data is to ascertain patterns in responses that correspond to a pre-defined set of causes of death.  Typically the nature of such patterns is not known \emph{a priori} and measurements are subject to multiple types of measurement error, discussed below.
   
Multiple methods have been proposed to automate the assignment of cause of death from VA data.  \rev{ The Institute for Health Metrics and Evaluation (IHME) has proposed a number of methods~\cite[for example: ][]{flax, james, murray}.  Many of these methods build on earlier work by King and Lu~\citep{king1, king2}.}   Altogether, this work has explored a variety of underlying statistical frameworks, although all are similar in their reliance on a so-called ``gold standard'' -- a database consisting of a large number of deaths for which the cause has been certified by medical professionals and is considered reliable.  Assuming the gold standard deaths are accurately labeled, methods in this class use information about the relationship between causes and symptoms from the gold standard deaths to infer causes for new deaths.  

Gold standard databases of this type are difficult and very expensive to create, and consequently most health researchers and policy makers do not have access to a good cause of death gold standard. Further it is often difficult to justify devoting medical professionals' time to performing autopsies and chart reviews for deceased patients in situations with limited resources.  Given these constraints, deaths included in a gold standard database are typically in-hospital deaths.  In most of the places where VA is needed, many or most deaths occur in the home and are not comparable to in-hospital deaths.  Further, the prevalence of disease changes dramatically through time and by region. In order to accurately reflect the relationship between VA data and causes of death, the gold standard would need to contain deaths from all regions through time; something that no existing gold standard does.

Recognizing the near impossibility of obtaining consistent gold standard databases that cover both time and space, we focus on developing a method to infer cause of death using VA data that does not require a gold standard.  A method developed by Peter Byass known as \textit{InterVA}~\citep{Byass:2012} has been used extensively, including by both the ALPHA and INDEPTH networks of HDSS sites, and is supported by the WHO.  Rather than using gold standard deaths to inform the relationship between signs and symptoms and causes of death, InterVA uses information obtained from physicians in the form of ranked lists of signs and symptoms associated with each cause of death.  

\rev{In this paper, we present a statistical method for assigning individual causes of death and population cause of death distributions using VA surveys in contexts where gold standard data are not available.  In the remainder of this section we describe the current practice and three critical limitations.  In Section~\ref{sec:insilico}  we propose a statistical model that addresses these three challenges and provides a flexible probabilistic framework that can incorporate the multiple types of data that are available in practice.  Section~\ref{sec:theresults} is the results section.  Section~\ref{sec:phmrc} presents results and comparison between the proposed method and existing alternatives using a gold standard dataset.  \revii{Section~\ref{sec:res} presents results from our method using data from two HDSS sites where no gold standard deaths are available.}  Section~\ref{sec:phys} describes how we incorporate physician assignment of cause of death.  Although these data integrate naturally into our method, we present them in a separate section because they are only relevant when physician-coded causes are available.  We end with Section~\ref{sec:concl} which provides a discussion of remaining limitations of VA data and proposes new directions for inquiry.  } \revii{Since many of the terms used in this paper are specific to the data and domain we discuss, we include a list of common acronyms in Table~\ref{tab:abbrevs}}

\begin{table}[htp]
\caption{List of common global health abbreviations used in this paper.}
\begin{center}
\begin{tabular}{|c|c|}
\hline
Acronym & Meaning\\
\hline
CSMF & Cause-specific mortality fraction\\
COD & Cause of death\\
HDSS & Health $\&$ demographic surveillance system\\
PHMRC & Population Health Medical Research Consortium\\
VA & Verbal autopsy\\
WHO & World Health Organization\\
\hline
\end{tabular}
\end{center}
\label{tab:abbrevs}
\end{table}%

\subsection{InterVA and three issues}
\rev{Given that InterVA is supported by the WHO and uses only information that is readily available in a broad range of circumstances, including those where there are no ``gold standard'' deaths, we focus our comparison on InterVA.  When evaluating the performance of our method, we use multiple methods for comparison and provide detailed descriptions of the alternative methods in the Online Supplement.}  InterVA \citep{byass2012strengthening} distributes a death across a pre-defined set of causes using information describing the relationship between VA data (signs and symptoms) and causes provided by physicians in a structured format.  The complete details of the InterVA algorithm are not fully discussed in published work, but they can be accurately recovered by examining the source code for the InterVA algorithm \citep{byass2012persComm}.  

Consider data consisting of $n$ individuals with observed set $S_{i}$ of (binary) indicators of symptoms, $S_{i}=\{s_{i1}, s_{i2}, ..., s_{iS}\}$ and an indicator $y_{i}$ that denotes which one of $C$ causes was responsible for person $i$'s death.  Many deaths result from a complex combination of causes that is difficult to infer even under ideal circumstances.  We consider this simplification, however, for the sake of producing demographic estimates of the fraction of deaths by cause, rather than for generating assignments that reflect the complexity of clinical presentations.  The goal is to infer $p(y_{i}=c|S_{i})$ for each individual $i$ and $\pi_c$, the overall population cause specific mortality fraction (CSMF) for cause $c$, for all causes.  Using Bayes' rule, 
\begin{equation}
p(y_{i}=c|S_{i})=\frac{p(S_{i}|y_{i}=c)p(y_{i}=c)}{p(S_{i}|y_{i}=c)p(y_{i}=c)+p(S_{i}|y_{i}\neq c)p(y_{i}\neq c)}.
\label{eq:bayes}
\end{equation}  
InterVA obtains the numerator in~(\ref{eq:bayes}) using information from interviews with a group of expert physicians.  \rev{Many of the methods currently available, whether or not they use gold standard data, make use of this logic.  That is, they rely on external input about the propensity of observing a particular symptom given the death was due to a particular cause.  Some methods use Bayes' rule directly (\citet{king2} uses Bayes' rule on a set of randomly drawn causes, for example) while others apply different transformations to obtain a ``propensity'' for each cause given a set of symptoms (e.g.~\citet{james}).  The key distinction for InterVA is that, for each cause of death, a group of physicians with expertise in a specific local context provide a ``tendency'' of observing each sign/symptom, presented in Table~\ref{tab:conditionalPs}. } These tendencies are provided in a form similar to the familiar letter grade system used to indicate academic performance.  These ``letter grades'', the leftmost column in Table~\ref{tab:conditionalPs}, effectively rank signs/symptoms by the tendency of observing them in conjunction with a given cause of death.  These rankings are translated into probabilities to form a cause by symptom matrix, ${\bf P}_{s|c}$.  InterVA uses the translation given in Table~\ref{tab:conditionalPs}.  This transformation in Table~\ref{tab:conditionalPs} is arbitrary and, as we demonstrate through simulation studies in 
\rev{the Online Supplement}, influential.  Our proposed method uses only the ranking and infers probabilities as part of a hierarchical model.
\begin{table}[h!]
\center
\captionsetup{format=plain,font=normalsize,margin=5cm,justification=justified}
\caption{InterVA Conditional Probability Letter-Value Correspondences from \cite{byass2012strengthening}.}
\begin{tabular}{l l l}
\hline
Label & Value & Interpretation \\
\hline\hline
I & 1.0 & Always \\
A+ & 0.8 & Almost always \\
A & 0.5 & Common \\
A- & 0.2 & \\
B+ & 0.1 & Often \\
B & 0.05 & \\
B- & 0.02 & \\
C+ & 0.01 & Unusual \\
C & 0.005 & \\
C- & 0.002 & \\
D+ & 0.001 &vRare \\
D & 0.0005 & \\
D- & 0.0001 & \\
E & 0.00001 & Hardly ever \\
N & 0.0 & Never \\
\hline \hline
\end{tabular}
\label{tab:conditionalPs}
\end{table}
The probability $p(S_{i}|y_{i}=c)$ in~(\ref{eq:bayes}) is the joint probability of an individual experiencing a set of symptoms (e.g. experiencing a fever and vomiting, but not wasting).  It is impractical to ask physicians about each combination of the hundreds of symptoms a person may experience.  Apart from being impossibly time-consuming, many combinations are indistinguishable because most symptoms are irrelevant for any particular cause.  InterVA addresses this by approximating the joint distribution of symptoms with the product of the marginal distribution for each symptom.  That is, $p(S_{i}|y_{i}=c)\approx \prod_{s=j}^{S} \{p(s_{ij}=1|y_{i}=c)\}^{s_{ij}} \{1-p(s_{ij}=1|y_{i}=c)\}^{1-s_{ij}}$.  This simplification is equivalent to assuming that the symptoms are conditionally independent given a cause, an assumption we believe is likely reasonable for many symptoms but discards valuable information in particular cases.  We discuss this assumption further in subsequent sections in the context of obtaining more information from physicians in the future. This challenge is not unique to our setting and also arises when using a gold standard dataset for a large set of symptoms.  Most of the methods described above that use gold standard data utilize a similar simplification, but they derive the necessary $p(S_{i}|y_{i}=c)$ empirically using the gold standard deaths (e.g. counting the fraction of deaths from cause $c$ that contain a given symptom).  

Three issues arise in the current implementation of InterVA.  First, although the motivation for InterVA arises through Bayes' rule, the implementation of the algorithm does not compute probabilities that are comparable across individuals.  InterVA defines $p(S_{i}|y_{i}=c)$ using only symptoms present for a given individual, that is $p(S_{i}|y_{i}=c) \triangleq \prod_{\{j : s_{ij}=1\}} p(s_{ij}=1|y_{i}=c)$.  The propensity used by InterVA to assign causes is based only on the presence of signs/symptoms, disregarding them entirely when they are absent:
\begin{equation}
p(y_{i} = c | S_{i}\in\{j : s_{ij}=1\})=\frac{p(y_{i} = c) \prod_{\{j : s_{ij}=1\}} p(s_{ij}=1 | y_i=c)} {\sum_{c=1}^{C} \left( \prod_{\{j : s_{ij}=1\}} p(s_{ij} | y_i=c) p(y_{i} = c) \right)}. 
\label{eq:present}
\end{equation}
The expression in~(\ref{eq:present}) ignores a substantial portion of the data, much of which could be beneficial in assigning a cause of death.  Knowing that a person had a recent negative HIV test could help differentiate between HIV/AIDS and tuberculosis, for example.  Using~(\ref{eq:present}) also means that the definition of the propensity used to classify the death depends on the number of positive responses.  If an individual reports a symptom, then InterVA computes the propensity of dying from a given cause conditional on that symptom.  In contrast, if the respondent does not report a symptom, InterVA marginalizes over that symptom.  Consider as an example the case where there are two symptoms $s_{i1}$ and $s_{i2}$.  If a respondent reports the decedent experienced both, then InterVA assigns the propensity of cause $c$ as $p(y_{i}=c | s_{i1}=1, s_{i2}=1)$.  If the respondent only reports symptom 1, the propensity is $p(y_{i}=c \cap s_{i2}=1 | s_{i1}=1)+p(y_{i}=c \cap s_{i2}=0 | s_{i1}=1)$.  These two measures represent fundamentally different quantities, so it is not possible to compare the propensity of a given cause across individuals. 

Second, because the output of InterVA has a different meaning for each individual, it is impossible to construct valid measures of uncertainty for InterVA.  We expect that even under the best circumstances there is variation in individuals' presentation of symptoms for a given cause.  In the context of VAs this variation is compounded by the added variability that arises from individuals' ability to recollect and correctly identify signs/symptoms.  
Linguistic and ethnographic work to standardize the VA interview process could control and help quantify these biases, though it is not possible to eliminate them completely.  Without a probabilistic framework, we cannot adjust the model for these sources of variation or provide results with appropriate uncertainty intervals.  This issue arises in constructing both individual cause assignments and population CSMFs.  The current procedure for computing CSMFs \rev{using InterVA} aggregates individual cause assignments to form CSMFs \citep{Byass:2012}.  This procedure does not account for variability in the reliability of the individual cause assignments, meaning that the same amount of information goes into the CSMF whether the individual cause assignment is very certain or little more than random guessing.  

Third, the InterVA algorithm does not incorporate other potentially informative sources of information.  VAs are carried out in a wide range of contexts with varying resources and availability of additional data.  For example, while true ``gold standard'' data are rarely available, many organizations already invest substantial resources in having physicians review at least a fraction of VAs and assign a cause based on their clinical expertise.  Physicians reviewing VAs are able to assess the importance of multiple co-occurring symptoms in ways that are not possible with current algorithmic approaches, and because of that, physician-assigned causes are a potentially valuable source of information.

In this paper, we develop a new method for estimating population CSMFs and individual cause assignments, \textit{InSilicoVA}, that addresses the three issues described above. Critically, the method is \emph{modular}.  At the core of the method is a general probabilistic framework.  On top of this flexible framework we can incorporate multiple types of information, depending on what is available in a particular context.  In the case of physician coded VAs, for example, we propose a method that incorporates physician expertise while also adjusting for biases that arise from their different clinical experiences.

\section{InSilicoVA}
\label{sec:insilico}
This section presents a hierarchical model for cause-of-death assignment, known as InSilicoVA.  This model addresses the three issues that currently limit the effectiveness of InterVA and provides a flexible base that incorporates multiple sources of uncertainty. \rev{We intend this method for use in situations where access to ``gold standard'' data is not possible and, as such, there are no labeled outcomes and we cannot leverage a traditional supervised learning framework.}  Section~\ref{sec:mainmodel} presents our modeling framework.  We then present the sampling algorithm in Section~\ref{sec:sampling}.

\subsection{Modeling framework}
\label{sec:mainmodel}
This section presents the InSilicoVA model, a hierarchical Bayesian framework for inferring individual cause of death and population cause distributions.  A key feature of the InSilicoVA framework is sharing information between inferred individual causes and population cause distributions.  As in the previous section, let $y_{i}=\{1,...,C\}$ be the cause of death indicator for a given individual $i$ and the vector $S_{i}=\{s_{i1}, s_{i2}, ..., s_{iS}\}$ be signs/symptoms.  We begin by considering the case where we have only VA survey data and will address the case with physician coding subsequently.  We typically have two pieces of information: (i) an individual's signs/symptoms, $s_{ij}$ and (ii) a matrix of conditional probabilities,  ${\bf P}_{s|c}$.  The ${\bf P}_{s|c}$ matrix describes a ranking of signs/symptoms given a particular cause.  


We begin by assuming that individuals report symptoms as independent draws from a Bernoulli distribution given a particular cause of death $c$.
That is, 
\[s_{ij}|y_{i}=c\sim\mbox{Bernoulli}(P(s_{ij}|y_{i}=c))\]
where $P(s_{ij}|y_{i}=c)$ are the elements of ${\bf P}_{s|c}$ corresponding to the given cause.  The assumption that symptoms are independent is likely violated, in some cases even conditional on the cause of death.  Existing techniques for eliciting the ${\bf P}_{s|c}$ matrix do not provide information about the association between two (or more) signs/symptoms occurring together for each cause, however, making it impossible to estimate these associations.  
%
Since $y_{i}$ is not observed for any individual, we treat it as a random variable.  Specifically, 
\[y_{i}|\pi_{1},...,\pi_{C}\sim\mbox{Multinomial}(\pi_{1},...,\pi_{C})\]
where $\pi_{1},...,\pi_{C}$ are the population cause-specific mortality fractions. 

Without gold standard data, we rely on the ${\bf P}_{s|c}$ matrix to understand the likelihood of a symptom profile given a particular cause.  In practice physicians provide only a ranking of likely signs/symptoms given a particular cause.  Rather than arbitrarily assigning probabilities to each sign/symptom in a particular ordering, as in Table~\ref{tab:conditionalPs}, we learn those probabilities.  
We could model each element of ${\bf P}_{s|c}$ using this expert information to ensure that, within each cause, symptoms with higher labels in Table~\ref{tab:conditionalPs} have higher probability.  Since many symptoms are uncommon, this strategy would require estimating multiple probabilities with very weak (or no signal) in the data.  Instead we estimate a probability for every letter grade in Table~\ref{tab:conditionalPs}.  This strategy requires estimating substantially fewer parameters and imposes a uniform scale across conditions.  Entries in the ${\bf P}_{s|c}$ matrix are not individual specific; therefore, we drop the $i$ indicator and refer to a particular symptom $s_{j}$ and entries in ${\bf P}_{s|c}$ as $p(s_{j}|y=c)$. Following~\citet{twl07}, we \rev{re-parameterize ${\bf P}_{s|c}$ as ${\bf P}_{L(s|c)}$, where $L(s|c)$ indicates the letter ranking of the $(s, c)$-th cell in the ${\bf P}_{s|c}$ matrix  based on the expert opinion provided by physicians in Table~\ref{tab:conditionalPs}.} We then give each entry in ${\bf P}_{L(s|c)}$ a truncated Beta prior:
\rev{\[
P_{L(s|c)}\sim \mbox{Beta}(\alpha_{s|c}, M - \alpha_{s|c}) \;\;\; \mbox{and} \;\;\;
P_{L(s|c)} \in (P_{L(s|c)-1},  P_{L(s|c)+1}), 
\]}
where $M$ and $\alpha_{s|c}$ are prior hyperparameters and are chosen so that ${\alpha_{s|c}}/{M}$ gives the desired prior mean for ${P}_{L(s|c)}$, 
\rev{and the constraint represents the truncation imposed by the order of the ranked probabilities. For simplicity, we use $\mathds{1}_{s|c}$ to represent an indicator of the interval where $P_{L(s|c)}$ is defined.
That is, $\mathds{1}_{s|c}$ defines the portion of a beta distribution between the symptoms with the next largest and next smallest probabilities, 
\[P_{L(s|c)}\sim\mathds{1}_{s|c}\mbox{Beta}(\alpha_{s|c}, M - \alpha_{s|c}).\]}
This strategy uses only the ranking (encoded through the letters in the table) and does not make use of arbitrarily assigned numeric values, as in InterVA.  Our strategy imposes a strict ordering over the entries of ${\bf P}_{s|c}$.  We could instead use a stochastic ordering by eliminating $\mathds{1}_{s|c}$ in the above expression.  Defining the size of $\alpha_{s|c}$ in an order consistent with the expert opinion in Table~\ref{tab:conditionalPs} would encourage, but not require, the elements of ${\bf P}_{s|c}$ to be consistent with expert rankings.  We find this approach appealing conceptually, but not compatible with the current strategy for eliciting expert opinion.  In particular, there are likely entries in ${\bf P}_{s|c}$ that are difficult for experts to distinguish.  In these cases it would be appealing to allow the method to infer which of these close probabilities is actually larger.  Current strategies for obtaining ${\bf P}_{s|c}$ from experts, however, do not offer experts the opportunity to report uncertainty, making it difficult to appropriately assign uncertainty in the prior distribution.     

We turn now to the prior distribution over population CSMF's, $\pi_{1},...,\pi_{C}$.  Placing a Dirichlet prior on the vector of population CSMF probabilities would be computationally efficient because of Dirichlet-Multinomial conjugacy.  However in our experience it is difficult to explain to practitioners and public health officials the intuition behind the Dirichlet hyperparameter.  Moreover, in many cases we can obtain a reasonably informed prior about the CSMF distribution from local public health officials.  Thus, we opt for an over-parameterized normal prior~\citep{gelman1996physiological} on the population CSMFs.  This prior representation does not enjoy the same benefits of conjugacy but is more interpretable and facilitates including prior knowledge about the relative sizes of CSMFs.  Specifically we model $\pi_{c}= \exp{\theta_{c}}/\sum_{c} \exp{\theta_{c}}$ where each $\theta_{c}$ has an independent Gaussian distribution with mean $\mu$ and variance $\sigma^{2}$.  We put diffuse uniform priors on $\mu$ and $\sigma^{2}$.  To see how this facilitates interpretability, consider a case where more external data exists for communicable compared to non-communicable diseases.  Then, the prior variance can be separated for communicable and non-communicable diseases to represent the different amounts of prior information.  
%
The model formulation described above yields the following posterior:
\begin{align*}
\Pr(\vec{y}, \mathbf{P}_{s|c}, \vec{\pi}, \mu, \sigma, \alpha |S_1,...,S_n) \propto& \prod_{i=1}^{n} \Pr({y_i}| \vec{\pi}) \prod_{j=1}^{S}\Pr(S_i|{y_i}, \mathbf{P}_{s|c}) \\ 
&\times\prod_{k=1}^{C}\prod_{j=1}^{S} \Pr({p}_{s_j|c_k}|\alpha) \times \prod_{k=1}^{C}\Pr(\pi_k| \mu, \sigma)\\
=& \prod_{i=1}^{n} \mbox{Categorical}(\vec{\pi})\prod_{j=1}^{S}\mbox{Bernoulli}(\mathbf{P}_{s|c}) \\ 
&\times\prod_{k=1}^{C}\prod_{j=1}^{S}  \mathds{1}_{s|c} \mbox{Beta}(\alpha_{s|c},\alpha_{s|c}-S)  \times \prod_{k=1}^{C}\mbox{Normal}(\mu_k, \sigma_k^2).
\end{align*}

To contextualize our work, we can relate it to Latent Dirichlet Allocation (LDA) and other text mining approaches to finding relationships between binary features.  To compare InSilicoVA to LDA, consider CSMFs as topics, conditions as words, and cases as documents.  InSilicoVA and LDA are similar in that we may consider each death as resulting from a combination of causes, just as LDA considers each document to be made up of a combination of topics.  Further, each cause in InSilicoVA is associated with a particular set of observed conditions, while in LDA each topic is associated with certain words.  The methods differ, however, in their treatment of topics (causes) and use of external information in assigning words (conditions) with documents (deaths).  Unlike LDA where topics are learned from patterns in the data, InSilicoVA is explicitly interested in inferring the distribution of a pre-defined set of causes of death.  InsilicoVA also relies on external information, namely the matrix of conditional probabilities ${\bf P}_{s|c}$ to associate symptoms with a given cause.  Statistically, this amounts to estimating a distribution of causes across all deaths, then using the matrix of conditional probabilities to infer the likely cause for each death, given a set of symptoms.  This means that each death arises as a mixture over causes, but inference about this distribution depends on both the pattern of observed signs/symptoms and the matrix of conditional probabilities.  In LDA, each document has a distribution over topics that is learned only from patterns of co-appearance between words.  We also note that the prior structure differs significantly from LDA to accomplish the distinct goals of VA.

\subsection{Sampling from the posterior}
\label{sec:sampling}


This section provides the details of our sampling algorithm.  We evaluated this algorithm through a series of simulation and parameter recovery experiments.  Additional details regarding this process are in the Online Supplement.  \rev{All codes are written in {\bf R}~\citep{rcitation}, with heavy computation done through calls to Java using {\bf rJava}~\citep{rJava}.  We have prepared an {\bf R} package which we have included with our resubmission and will submit the package to CRAN prior to publication.}  

\subsubsection{Metropolis-within-Gibbs algorithm}
The posterior in the previous section is not available in closed form.  We obtain posterior samples using Markov-chain Monte Carlo, specifically the Metropolis-within-Gibbs algorithm described below.  
%
We first give an overview of the entire procedure and then explain the truncated beta updating step in detail.  
Given suitable initialization values, the sampling algorithm proceeds:
\begin{enumerate}
\item Sample $P_{s|c}$ from truncated beta step described in the following section.

\item Generate $Y$ values using the Naive Bayes Classifier, that is for person $i$
\[y_{i}|\vec{\pi} ,S \sim \mbox{Categorical}\left(p^{(NB)}_{1i},p^{(NB)}_{2i},...p^{(NB)}_{Ci}\right)\] where 
\[p^{(NB)}_{ci}=\frac{ \pi_{c} \prod_{j=1}^{S}(P(s_{ij}=1|y_{i}=c))^{s_{ij}}(1-P(s_{ij}=1|y_{i}=c))^{1-s_{ij}}}{\sum_{c}\pi_{c}\prod_{j=1}^{S}(P(s_{ij}=1|y_{i}=c))^{s_{ij}}(1-P(s_{ij}=1|y_{i}=c))^{1-s_{ij}}}\]
\item Update $\vec{\pi}$
	\begin{enumerate}

		\item Sample $\mu$
			\[ \mu \sim N\left(\frac{1}{C}\sum_{k=1}^C \theta_k, \,  \frac{\sigma^{2}}{C} \right)
			\]
		\item Sample $\sigma^2$
			\[ \sigma^2 \sim \mbox{Inv-}\chi^2\left(C-1, \, \frac{1}{C}\sum_{i=1}^n (\theta_k - \mu)^2\right)
			\]
		\item Sample $\vec{\theta} $
			\[ \vec{\theta} \propto \mbox{Multinomial}(N, \pi_k) \cdot (N(\mu, \sigma^2))^C 
			\]
			This needs to be done using a Metropolis Hastings step: for $k$ in 1 to C, 
			\begin{itemize}
			\item Sample $U \sim \mbox{Uniform}(0, 1)$
			\item Sample $\vec{\theta}^{*} \sim N(\vec{\theta}, \sigma^{*})$, 
			\item If $U \leq \frac{\prod_{k=1}^C(\pi^{*}_k)^{n_k} \exp{\frac{-(\theta^{*}_k - \mu)^2}{2\sigma^2}}}{\prod_{k=1}^C(\pi_k)^{n_k} \exp{\frac{-(\theta_k - \mu)^2}{2\sigma^2}}}$, then update $\theta_k$ by $\theta^{*}_k$.
            \end{itemize}						
	\end{enumerate}
\end{enumerate}
We find computation time to be reasonable even for datasets with $\sim 10^5$ deaths.   We provide additional details about assessing convergence in the results section and Online Supplement.

\subsubsection{Truncated beta step}

As described in the previous section, our goal is to estimate probabilities in $\mathbf{P}_{s|c}$ for each ranking given by experts (the letters in Table~\ref{tab:conditionalPs}).  We denote the levels of $\Pr(s|c)$ as $L(s|c)$ and, assuming the prior from the previous section,
sample the full conditional probabilities under the assumption that all entries with the same level in $\mathbf{P}(s|c)$ still share the same value.  Denoting the probability for a given ranking or tier as $P^{t}(s_i|c_j)$, the full constraints become:
\begin{align*} 
P^{t}(s_i|c_j) &= P^{t}(s_k|c_{j'}), \; \forall k\; s.t. \; L(s_i | c_j) = L(s_j | c_{j'}) \\
P^{t}(s_i|c_j) &< P^{t}(s_k|c_{j'}), \; \forall k \; s.t. \; L(s_i | c_j) < L(s_k | c_{j'}) \\
P^{t}(s_i|c_j) &> P^{t-1}(s_k|c_{j'}), \; \forall k \; s.t. \; L(s_i|c_j) > L(s_k|c_{j'}).
\end{align*}
The full conditionals are then truncated beta distributions with these constraints, defined as: \begin{align*}
\Pr(s_j|c_k, \mathbf{S}, \vec{y}) &= P_{L(s_j|c_k) | \mathbf{S}, \vec{y}}  \sim \\
\mathds{1}_{s|c} \mbox{Beta}\Bigg(\alpha_{L_{s_j|c_k}} + \sum_{\substack{j',k' : \\L(s_{j'}|c_{k'}) = L(s_j|c_k)}}\# \{S_{j'}|c_{k'}\} \ & , \
M - \alpha_{L_{s_j|c_k}} + \sum_{\substack{j',k' : \\L(s_{j'}|c_{k'}) = L(s_j|c_k)}}( n_{c_k'}- \# \{S_{j'}|c_{k'}\})\Bigg),
\end{align*}
where $M$ and $\alpha_{L_{s_j|c_k}}$ are hyperparameters and $\vec{y}$ is the vector of causes at a given iteration.  \rev{The $\mathbf{1}_{S|C}$ term defines an indicator function which we denote as short hand for $\mathbf{1}_{P_{L(s|c)} \in (P_{L(s|c)-1},  P_{L(s|c)+1}) }$.  That is, $\mathbf{1}_{S|C}$ denotes the indicator for whether the level for a particular $s|c$ falls between the upper and lower bounds of that level.}
We incorporate these full conditionals into the sampling framework above, updating the truncation at each iteration according to the current values of the relevant parameters. 

\section{Results}
\label{sec:theresults}
\rev{In this section we present results from our method in two contexts.  First, in Section~\ref{sec:phmrc} we compare our method to alternative approaches using a set of gold standard deaths.  For these results, we compared our method to both InterVA and to methods currently used in practice when gold standard deaths are available.  Then, in Section~\ref{sec:res} we implement our method using data from two health and surveillance sites.}

\rev{Along with the results presented in the remainder of this section, we also performed simulation studies to understand our method's range of performance.  Details and results of these simulations are presented in the Online Supplement.} 

\subsection{PHMRC gold standard data}
\label{sec:phmrc}
\rev{
We compare the performance of multiple methods using the Population Health Medical Research Consortium (PHMRC) dataset~\citep{murray2011population}.  The PHMRC dataset consists of about 7,000 deaths recorded in six sites across four countries (Andhra Pradesh, India; Bohol, Philippines; Dar es Salaam, Tanzania; Mexico City, Mexico; Pemba Island, Tanzania; and Uttar Pradesh, India).  Gold standard causes are assigned using a set of specific diagnostic criteria that use laboratory, pathology, and medical imaging findings.  All deaths occurred in a health facility.  For each death, a blinded verbal autopsy was also conducted. }

\rev{To evaluate the performance of InSilicoVA, we compared our proposed method to a number of alternative algorithmic and probabilistic methods.  Specifically, we examined Tariff~\citep{james}, the approach proposed by~\citet{king2}, the Simplified Symptom Pattern Method~\citep{murray} and InterVA.  We provide a short description of each comparison method along with complete details of our implementation in the Online Supplement.  The most realistic comparator to our method is InterVA, since InterVA is the only method currently available that does not require gold standard training data. For both InterVA and InSilicoVA, we use the training data to extract a matrix of ranks that used to estimate a ${\bf P}_{s|c}$ matrix.  In practice, these values would come from input provided by experts.}
 \revii{Through simulation studies provided in the Online Supplement, we show that the quality of inputs greatly influences performance of all methods.  
 With the PHMRC data we use the labeled deaths to generate inputs for all of the methods, separating the performance of the statistical model or algorithm from the quality of inputs.}  

\rev{We calculate the raw conditional probability matrix of each symptom given a cause, then used two approaches to construct the rank matrix from the empirical values required for InterVA and InSilicoVA.
First, we ranked the conditional probabilities using $15$ levels and a distribution matching the InterVA model. For example, if $a\%$ of the cells in the original InterVA ${\bf P}_{s|c}$ matrix are assigned the lowest level, we assign also $a\%$ of the cells in the empirical matrix to be that level. We assign the median value among these cells to be the default value for this level.  We refer to this first approach as the ``Quantile prior'' since it preserves the same distribution of probabilities in the InterVA matrix.  Second, we ranked the probabilities using the same levels in Table~\ref{tab:conditionalPs} by assigning each cell the letter grade in Table~\ref{tab:conditionalPs} with value closest to it.  We refer to this approach as the ``Default prior" since it uses the same translation between empirical probabilities and ranks as in InterVA.  
}

\rev{In examining the results, note that SSP does not provide CSMF estimates and~\citet{king2} does not provide individual cause assignments.  For the individual cause assignment we also evaluated performance using the Chance-Corrected Concordance metric proposed by~\citet{murray2014using}.  Results using this metric were substantively similar, with the exception that the Tariff method's performance decreases when viewed with this metric.  Complete results with chance-corrected concordance are presented in the Online Supplement.}

\revii{We evaluated out of sample performance in two ways.  Before considering these results, we note that InSilicoVA and InterVA use less information from the data than the comparison methods.  In each comparison, the methods designed for gold standard data have access to all available information in the training set, whereas InSilicoVA and InterVA only have access to the ranked likelihood of seeing a symptom for a given cause.}  

\begin{figure}[!h]
\centering
\includegraphics[width=.48\textwidth]{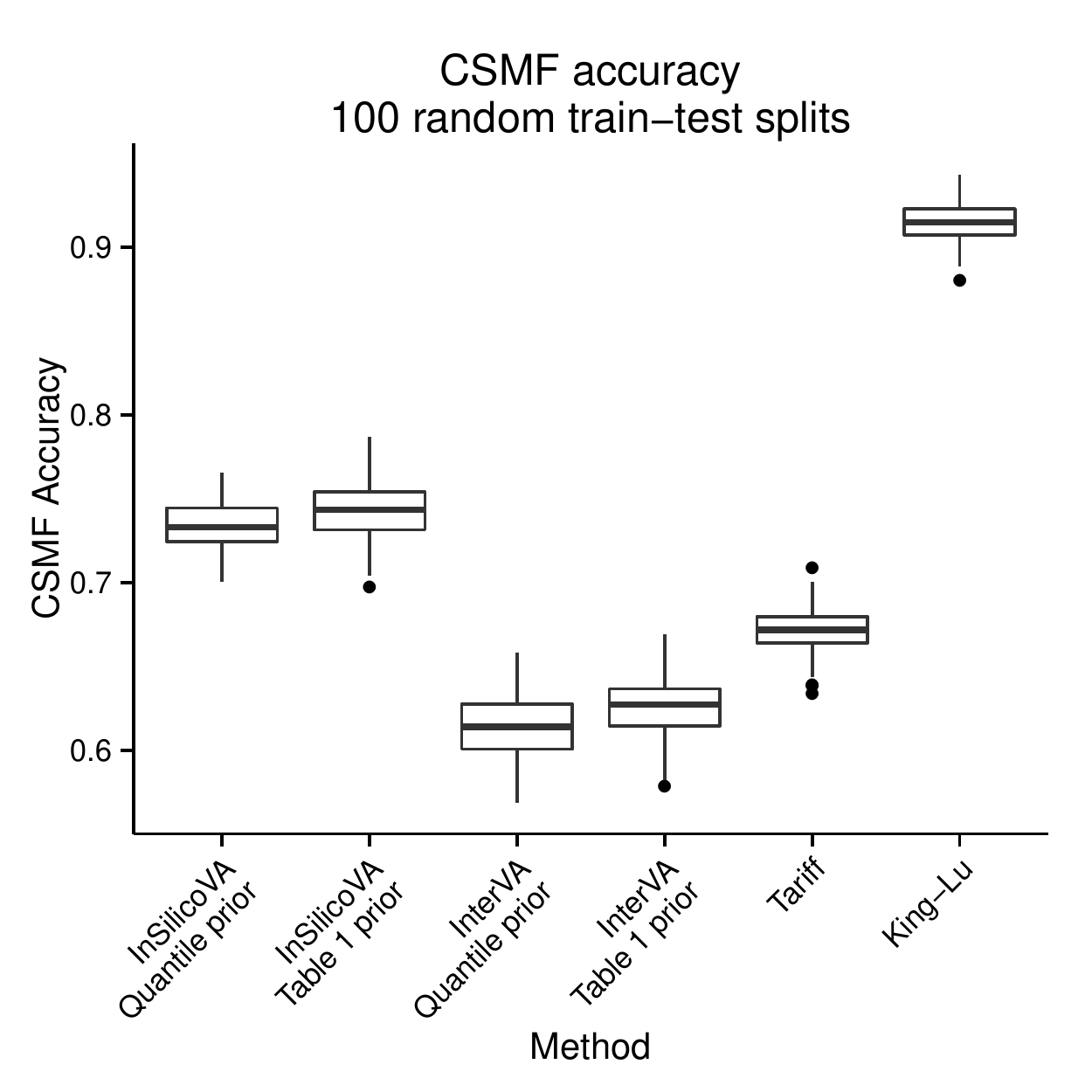}
\includegraphics[width=.48\textwidth]{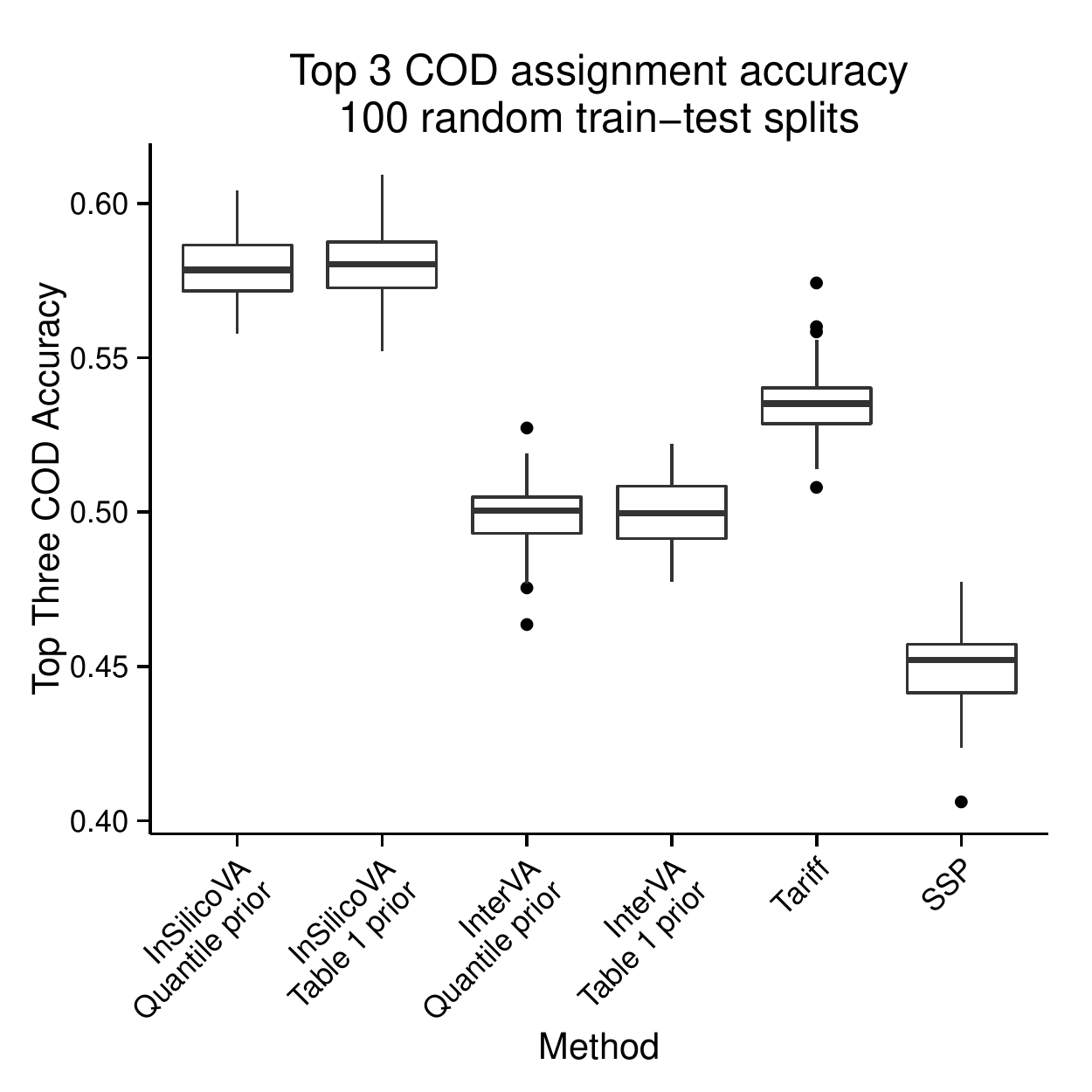}
\caption{\bf Comparison using random splits.}
\floatfoot{Results across 100 test-train splits using a randomly sampled 25\% of deaths as the test set.  InSilicoVA demonstrates good performance for both population cause distribution and individual assignment.  The~\citet{king2} method performs extremely well because the method is more likely to estimate the CSMF to be similar to the training CSMF, as expected using simple random samples to do test-train splits.}
\label{fig:random}
\end{figure}

\rev{In our first evaluation, we obtained training sets by simply sampling each death with equal probability.  This is the most straightforward approach to derive test-train splits and produces a CSMF distribution that matches the data.  Figure~\ref{fig:random} displays the results across 100 test-train splits.  A striking first result is the unparalleled performance of the~\citet{king2} method, which occurs because the~\citet{king2} method is more likely to estimate CSMFs that are similar to the training CSMFs, especially when the number of symptoms is large.  When the distribution in the training and testing sets are similar (as expected using simple random samples to do test-train splits)~\citet{king2} uses this information effectively.  This is deceiving, however, because in practice if one were able to achieve a simple random sample of deaths with reasonably assigned causes there would be little need for further methods.  Aside from this caveat, InSilicoVA displays superior performance in both CSMF distribution and individual cause assignments. The two ways of computing $P_{s|c}$ make little difference in the performance, indicating that the model can recover reasonable estimates of $P_{s|c}$ entries even when ranks are constructed in different ways.}  

\rev{For our second PHMRC evaluation study, we use the test/training split approach proposed by~\citet{murray2011population}.  Using this approach, the dataset is first randomly split into testing and training sets.  For the testing set, a cause of death distribution is simulated from a Dirichlet distribution at each iteration.  Cases from the testing set are then sampled with replacement with probability proportional to the simulated CSMF.  This approach allows the ``true'' cause of death distribution to change at each iteration, meaning  that results are not specific to a single cause of death distribution.  This approach also addresses the issue which arises in Figure~\ref{fig:random} where the method by~\citet{king2} has (unrealistically) nearly perfect performance. A downside of this re-sampling approach, however, is that the implementation proposed by~\citet{murray2011population} uses a diffuse Dirichlet distribution, meaning that the cause of death distribution is typically flat across causes.  In practice, the data we use display large variation in the cause of death distributions, with a small number of causes accounting for most of the deaths.}

\begin{figure}[!h]
\centering
\includegraphics[width=.48\textwidth]{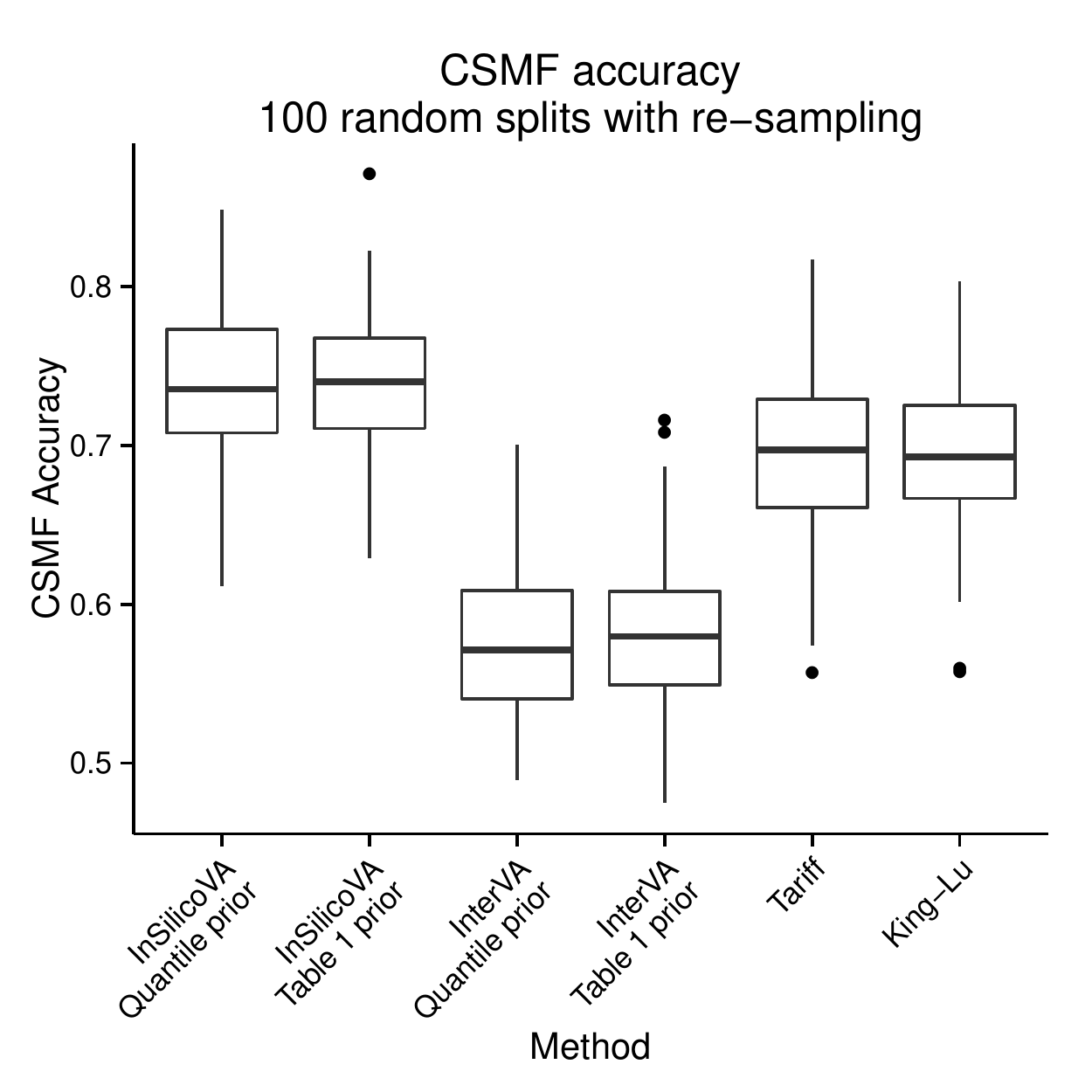}
\includegraphics[width=.48\textwidth]{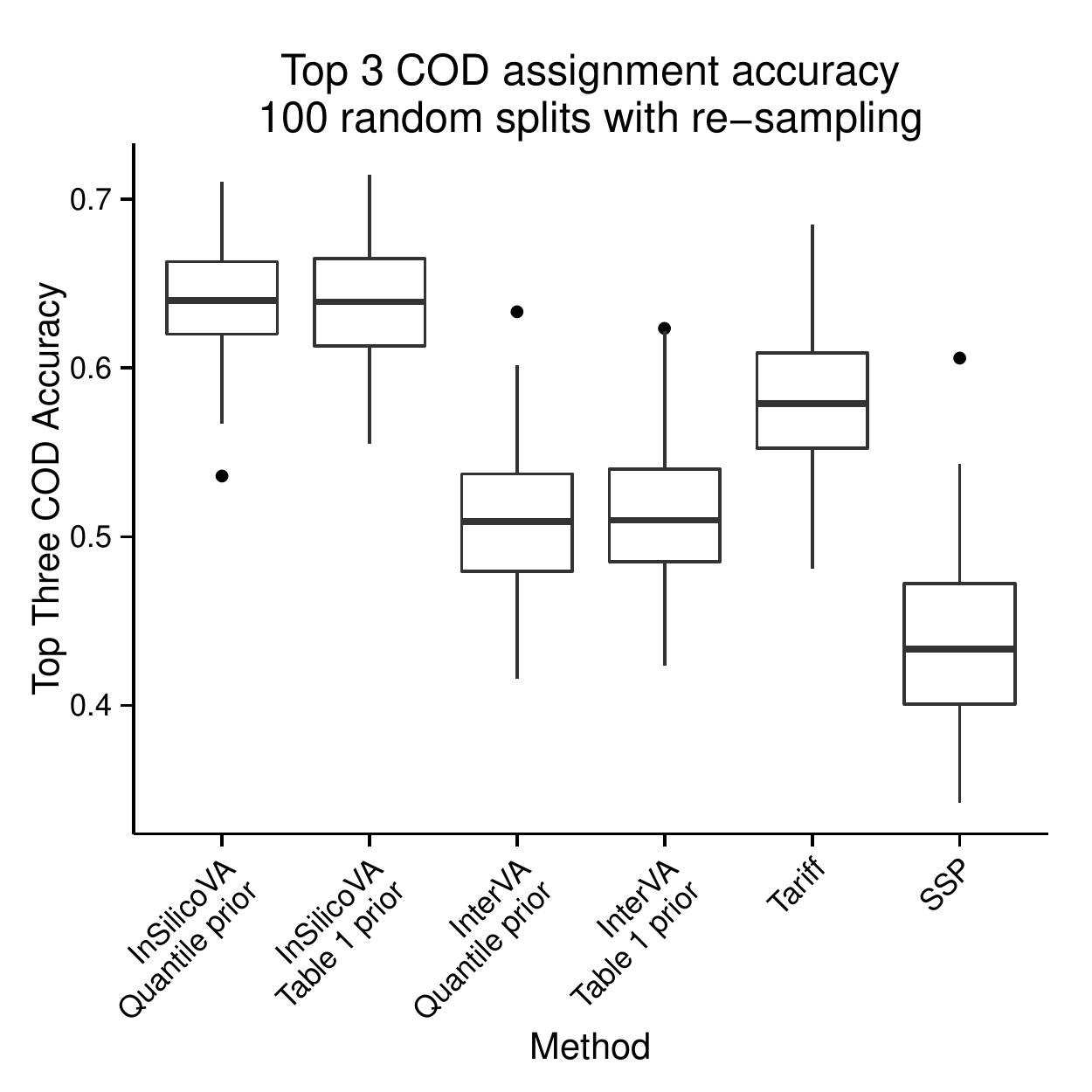}
\caption{\bf Comparison using random splits and re-sampled test set.}
\floatfoot{Performance over 100 simulations using the test-train splitting approach proposed by~\citet{murray2011population}.  The Dirichlet sampling approach used by the~\citet{murray2011population} method produces testing sets that do not necessarily have CSMF distributions similar to the training set, resulting in lower performance by the~\citet{king2} method.  InSilicoVA again demonstrates substantial performance improvements.}
\label{fig:dirichlet}
\end{figure}
\begin{figure}[!h]
\includegraphics[width=.49\textwidth]{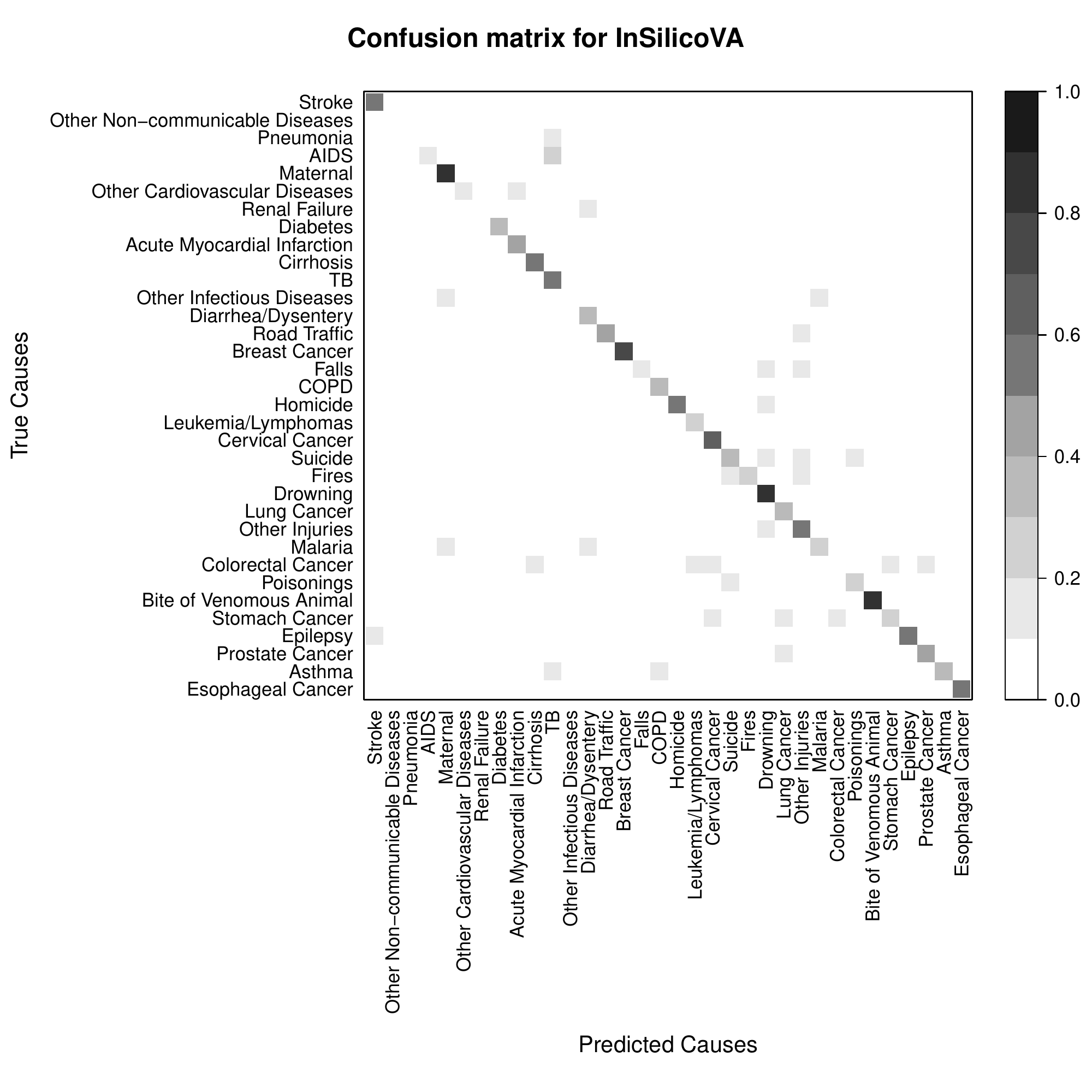}
\includegraphics[width=.49\textwidth]{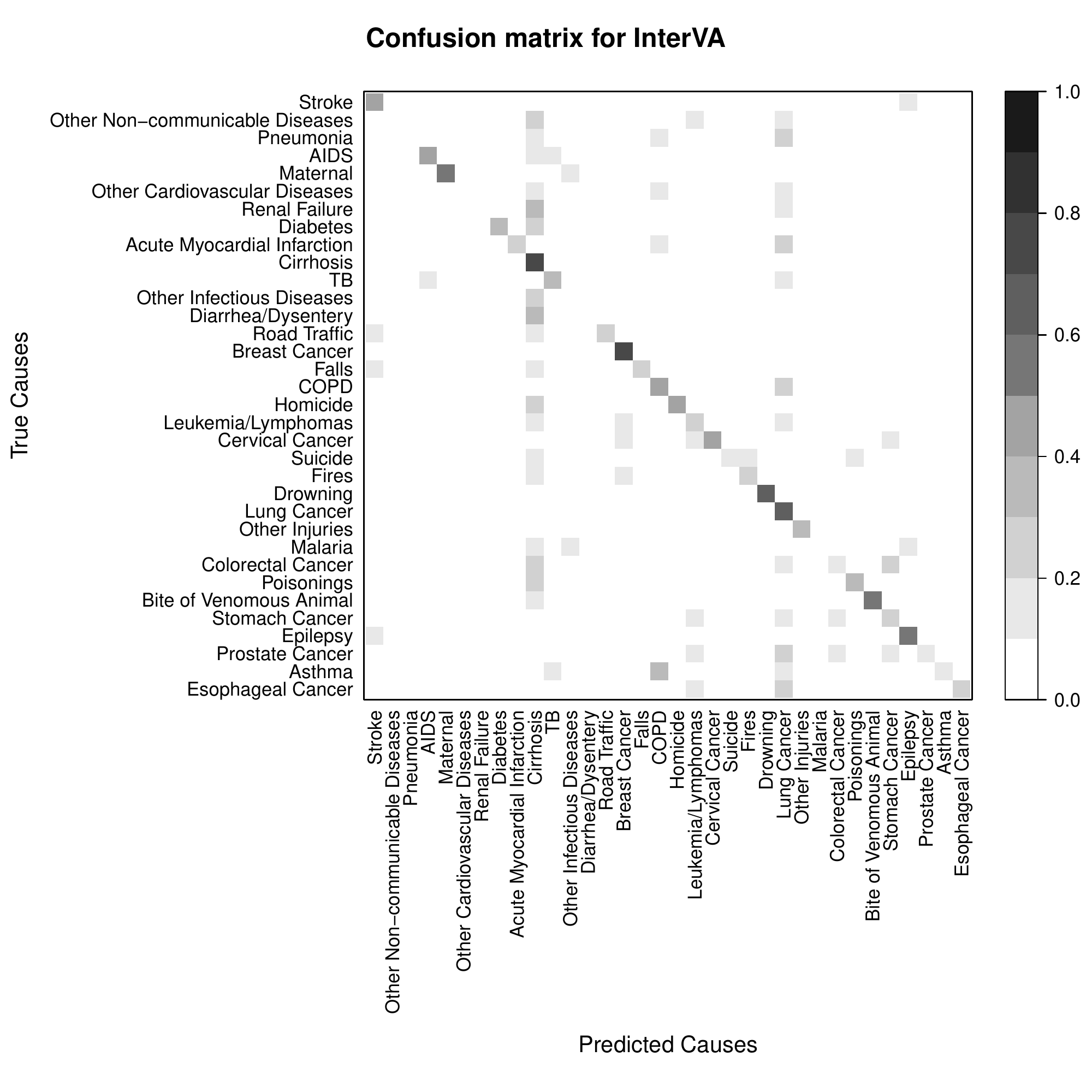}\\
\includegraphics[width=.49\textwidth]{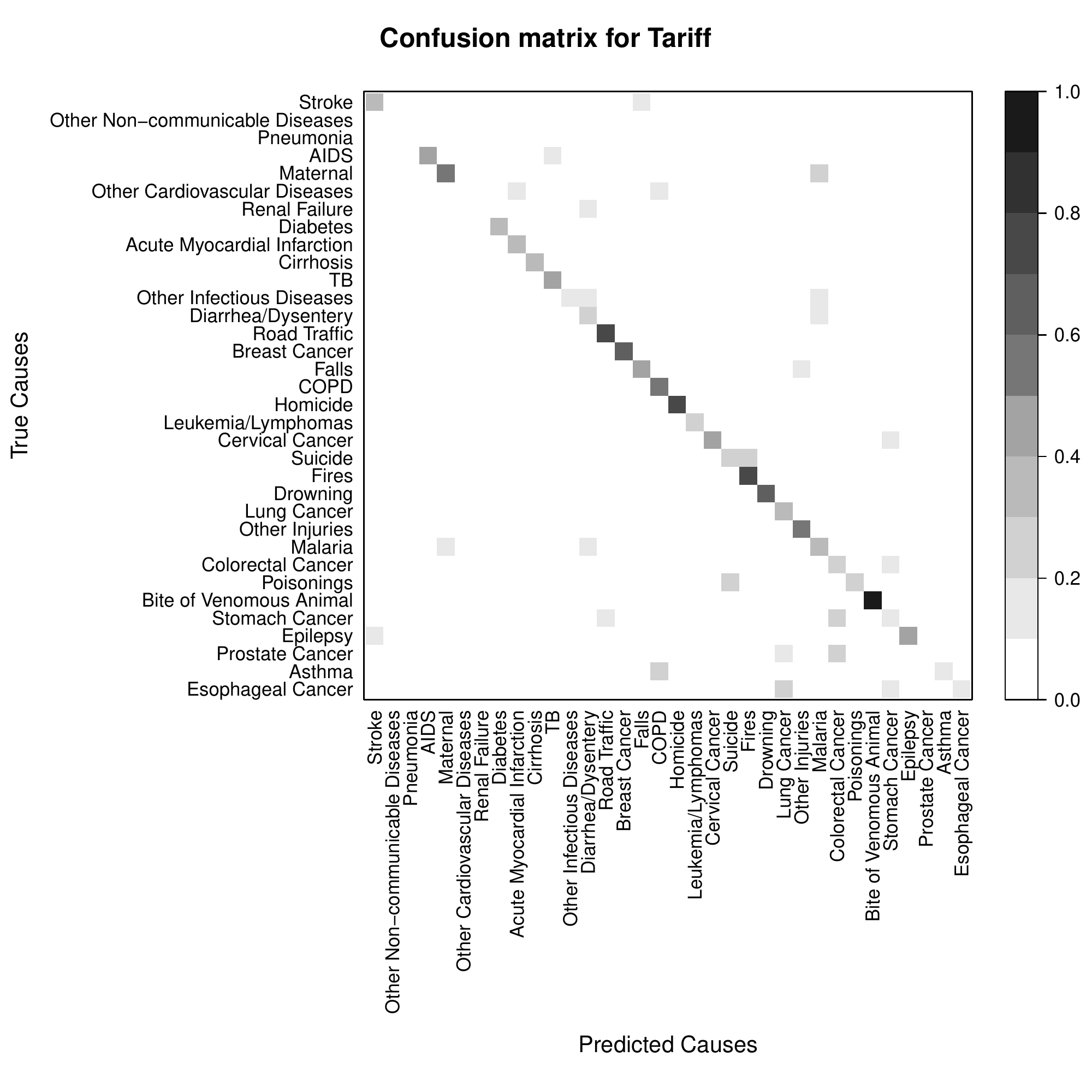}
\includegraphics[width=.49\textwidth]{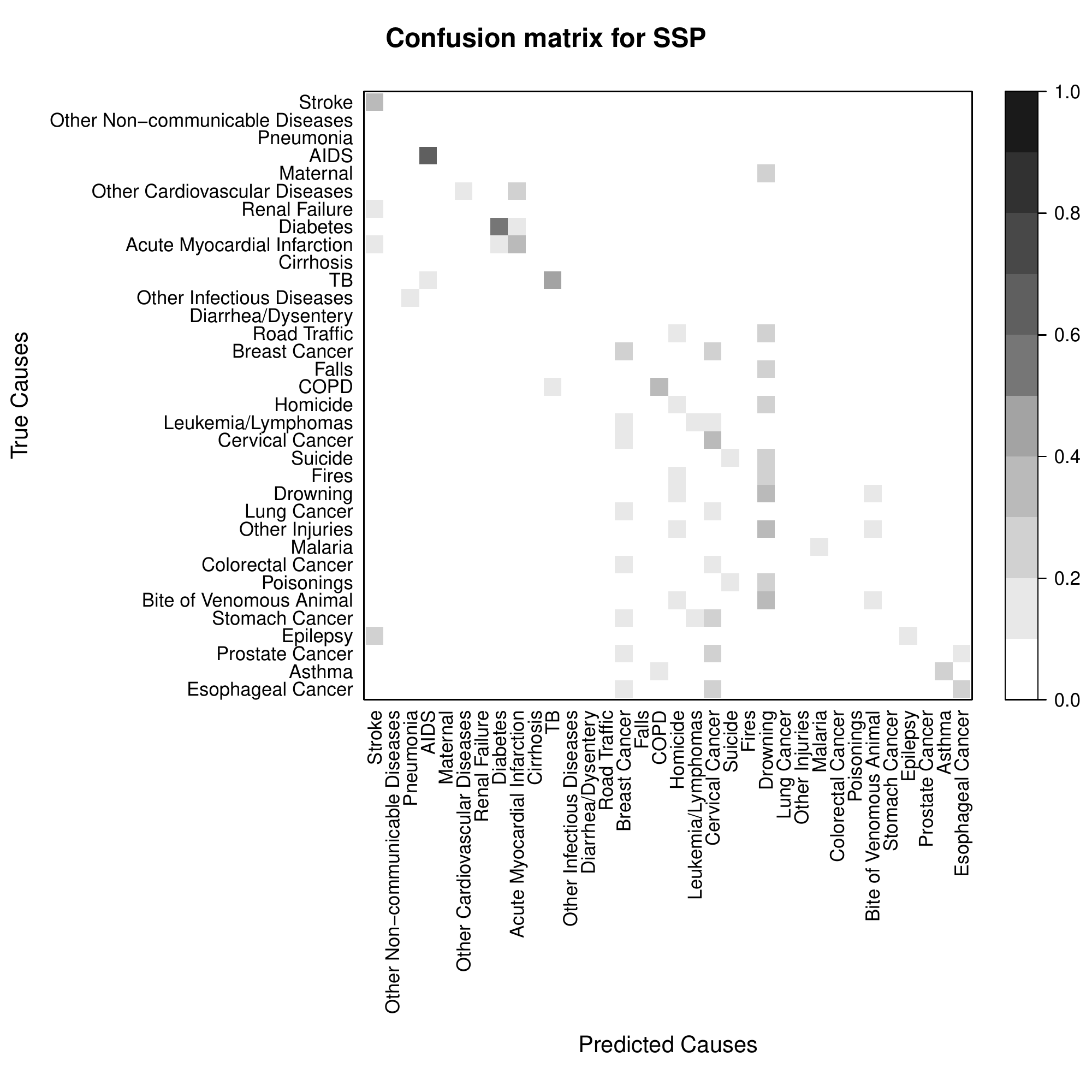}
\caption{\bf Confusion matrices.}
\floatfoot{Confusion matrix for four methods, with InSilicoVA and InterVA using the default prior.  Shading on the diagonal represents the fraction of correct classifications across 100 test-train splits using the Dirichlet resampling strategy; Off-diagional shading represents the fraction of misclassifcations.}
\label{fig:confusion}
\end{figure}

Figure~\ref{fig:dirichlet} displays the results for the Dirichlet assignment evaluation.  There is greater variability in performance among all methods compared to Figure~\ref{fig:random}, though InSilicoVA displays superior performance in both CSMF and individual cause assignment.  As expected, forcing the~\citet{king2} method to infer a cause distribution in the test set that is different from the training set produces a more fair comparison.  
\revii{To better understand how the methods perform on specific causes, Figure~\ref{fig:confusion} shows a confusion matrix for the Dirichlet resampling evaluation study.  The diagonals in the confusion matrices represent the number of times the method correctly identifies a cause $X$ across 100 simulations, divided by the number of times across the 100 test-train splits that cause $X$ was the true cause.  The off-diagonal elements give the fraction of times the method incorrectly identifies cause $X$ as the true cause across the test-train splits.  For InSilicoVA and InterVA we show results for the default (InterVA cutoffs) prior, though the results for the quantile prior were substantively similar.  We see the best performance from InSilicoVA and Tariff.  Note also that many of the cases involving misclassification in all of the methods occur where there are related causes.  HIV/AIDs and Tuberculosis (TB) are frequently misclassified, for example.  Many cancers are also difficult to differentiate from one-another, as we see off-diagional shading in the lower center and right of the plots.  We provide further results in the Online Supplement, including tables giving sensitivity and specificity for each method and cause.}   
%

\rev{Throughout this work, we claim that gold standard data are rarely available in practice.  One may question, however, why we cannot simply use the PHMRC data as a gold standard dataset in future applications.  To evaluate this question, we exploit the geographic variability within the PHMRC data.} 
\rev{We used the data from one of the PHMRC sites as the testing set, then used as the training set: (1) all the sites, (2) all other sites, (3) the same site and (4) one of the other sites.  The results are summarized in  \revii{Tables~\ref{tab:simsites1} and \ref{tab:simsites2}.  We show additional results in the Online Supplement that reveal performance for each site as training data individually.  In almost every case InSilicoVA outperforms all other methods in terms of both mean CSMF accuracy and mean COD assignment accuracy.} The only exception is when using one site as training and a different site as testing, where~\citet{king2} has slightly higher average CSMF accuracy.  We see a noticeable decrease in performance when training using a site that is different from the testing region.  The probabilistic structure underlying InSilicoVA and using only the conditional probabilities given in the ${\bf P}_{s|c}$ matrix mitigate this effect, though we still see a performance drop.  This indicates that both the distribution of causes of death \emph{and} that the association between symptoms and causes change based on geographic variability.  In contrast to the resources required to collect gold standard data, the ${\bf P}_{s|c}$ matrix used by InterVA and InSilicoVA can be collected cheaply and efficiently and tailored to the specific symptom/cause relationship in a particular geographic setting.}

 \begin{table}[!h]
 \centering
 \begin{tabular}{r|rrrr|rrrr}
   \hline
 Training& \multicolumn{2}{c}{All sites} & \multicolumn{2}{c}{All other sites}& \multicolumn{2}{c}{Same site}& \multicolumn{2}{c}{One other site} \\
  Testing& mean & sd & mean & sd & mean & sd &mean &sd\\ 
   \hline
 InSilicoVA - Quantile prior & \bf0.70 & 0.07 & 0.60 & 0.06 & 0.84 & 0.05 & 0.52 & 0.12 \\ 
   InSilicoVA - Default prior & 0.68 & 0.06 & \bf0.61 & 0.09 & \bf 0.85 & 0.05 & 0.52 & 0.12 \\ 
   InterVA - Quantile prior & 0.57 & 0.11 & 0.56 & 0.13 & 0.72 & 0.07 & 0.49 & 0.12 \\ 
   InterVA -  Default prior& 0.57 & 0.12 & 0.56 & 0.14 & 0.76 & 0.07 & 0.49 & 0.12 \\ 
   Tariff & 0.62 & 0.10 & 0.58 & 0.11 & 0.64 & 0.06 & 0.47 & 0.13 \\ 
   King-Lu & 0.64 & 0.11 & 0.60 & 0.13 & - & - & \bf 0.56 & 0.12 \\
    \hline
 \end{tabular}
 \caption{Mean and standard deviation of CSMF accuracies tested on each site using different training sets.}
 \label{tab:simsites1}
 \end{table}
 \begin{table}[!h]
 \centering
 \begin{tabular}{r|rrrr|rrrr}
   \hline
 Training& \multicolumn{2}{c}{All sites} & \multicolumn{2}{c}{All other sites}& \multicolumn{2}{c}{Same site}& \multicolumn{2}{c}{One other site} \\
  Testing& mean & sd & mean & sd & mean & sd &mean &sd\\ 
   \hline
 InSilicoVA - Quantile prior & \bf 0.64 & 0.04 &\bf 0.52 & 0.06 & \bf 0.82 & 0.09 & 0.41 & 0.10 \\ 
   InSilicoVA - Default prior & \bf 0.64 & 0.05 &\bf 0.52 & 0.04 & \bf 0.83 & 0.08 & \bf 0.42 & 0.09 \\ 
   InterVA - Quantile prior & 0.54 & 0.05 & 0.44 & 0.05 & 0.71 & 0.12 & 0.34 & 0.07 \\ 
   InterVA - Default prior & 0.54 & 0.06 & 0.43 & 0.06 & 0.77 & 0.10 & 0.25 & 0.11 \\ 
   Tariff & 0.54 & 0.08 & 0.48 & 0.07 & 0.60 & 0.06 & 0.31 & 0.10 \\ 
   SSP & 0.54 & 0.06 & 0.36 & 0.05 & 0.78 & 0.05 & 0.29 & 0.05 \\ 
    \hline
 \end{tabular}
 \caption{Mean and standard deviation of top 3 COD assignment accuracy accuracies tested on each site using different training sets.}
 \label{tab:simsites2}
 \end{table}


\subsection{HDSS sites}\label{sec:res}
In this section we present results comparing InSilicoVA and InterVA using VA data from two Health and Demographic Surveillance Sites (HDSS).  Section~\ref{sec:realStudy} provides background information to contextualize the diverse environments of the two sites, and Section~\ref{sec:results} presents the results.

\subsubsection{Background: Agincourt and Karonga sites}\label{sec:realStudy}
We apply both methods to VA data from two HDSS sites: the Agincourt health and socio-demographic surveillance system~\citep{kahn2012profile} and the Karonga health and demographic surveillance system~\citep{crampin2012profile}.  \revii{Both sites collect VA data as the primary means of understanding the distribution and dynamics of cause of death.  As is typically the case in circumstances where VAs are used, the VA data at these sites are not validated by physicians performing physical autopsies.  We do not, therefore, have gold standard data we can leverage for training methods.}  

The Agincourt site continuously monitors the population of about 31 villages located in the Bushbuckridge subdistrict of Ehlanzeni District, Mpumalanga Province in northeast South Africa. This is a rural population living in what was during Apartheid a black ``homeland,'' or Bantustan. The Agincourt HDSS was established in the early 1990s to guide the reorganization of South Africa's health system. Since then the site has functioned continuously and its purpose has evolved so that it now conducts health intervention trials and contributes to the formulation and evaluation of health policy. The population covered by the site is approximately 120,000 and vital events including deaths are updated annually.  VA interviews are conducted on every death occurring within the study population.  We use 9,875 adult deaths from Agincourt from people of both sexes from 1993 to the present.  

The Karonga site monitors a population of about 35,000 in northern Malawi near the port village of Chilumba.  The current system began with a baseline census from 2002--2004 and has maintained continuous demographic surveillance.  The Karonga site is actively involved in research on HIV, TB, and behavioral studies related to disease transmission. Similar to Agincourt, VA interviews are conducted on all deaths, and this work uses 1,469 adult deaths from Karonga that have occurred in people of both sexes from 2002 to the present.

\subsubsection{Results for HDSS sites}\label{sec:results}
We fit InSilicoVA to VA data from both Agincourt and Karonga.  We also fit InterVA using the physician-generated conditional probabilities $\mbf{P}_{s|c}$ as in Table ~\ref{tab:conditionalPs} and the same prior CSMFs $\vec{\pi}$ provided by the InterVA software \citep{2013interVA}.  We removed external causes (e.g., suicide, traffic accident, etc.) because deaths from external causes usually have very strong signal indicators and are usually less dependent on other symptoms.  \rev{For InterVA, we first removed the deaths clearly from external causes and use the R package \textit{InterVA4} \citep{li2014interva4} to obtain CSMF estimates.} For InSilicoVA, we ran three MCMC chains with different starting points. For each chain, we ran the sampler for $10^4$ iterations, discarded the first $5,000$,  and then thinned the chain using every twentieth iteration. \rev{We ran the model in R and it took about $15$ minutes for Agincourt data and less than $5$ minutes for Karonga using a standard desktop machine}.  Visual inspection suggested good mixing and we assessed convergence using Gelman-Rubin \citep{gelman1992inference} tests.  Complete details of our convergence checks are provided in the Online Supplement. 

\begin{figure}[htb]
\centering
\includegraphics[width = .49\textwidth]{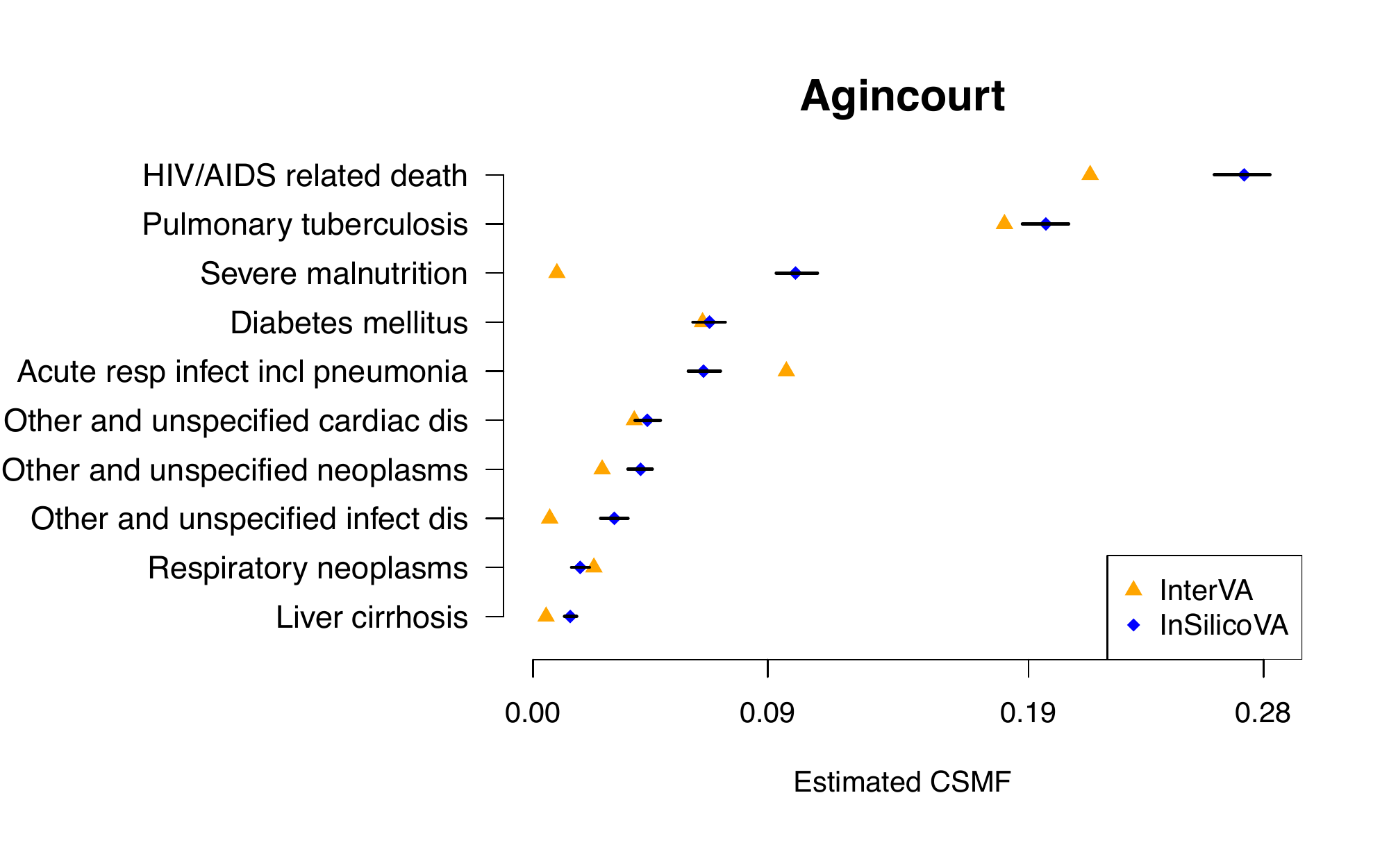}
\includegraphics[width = .49\textwidth]{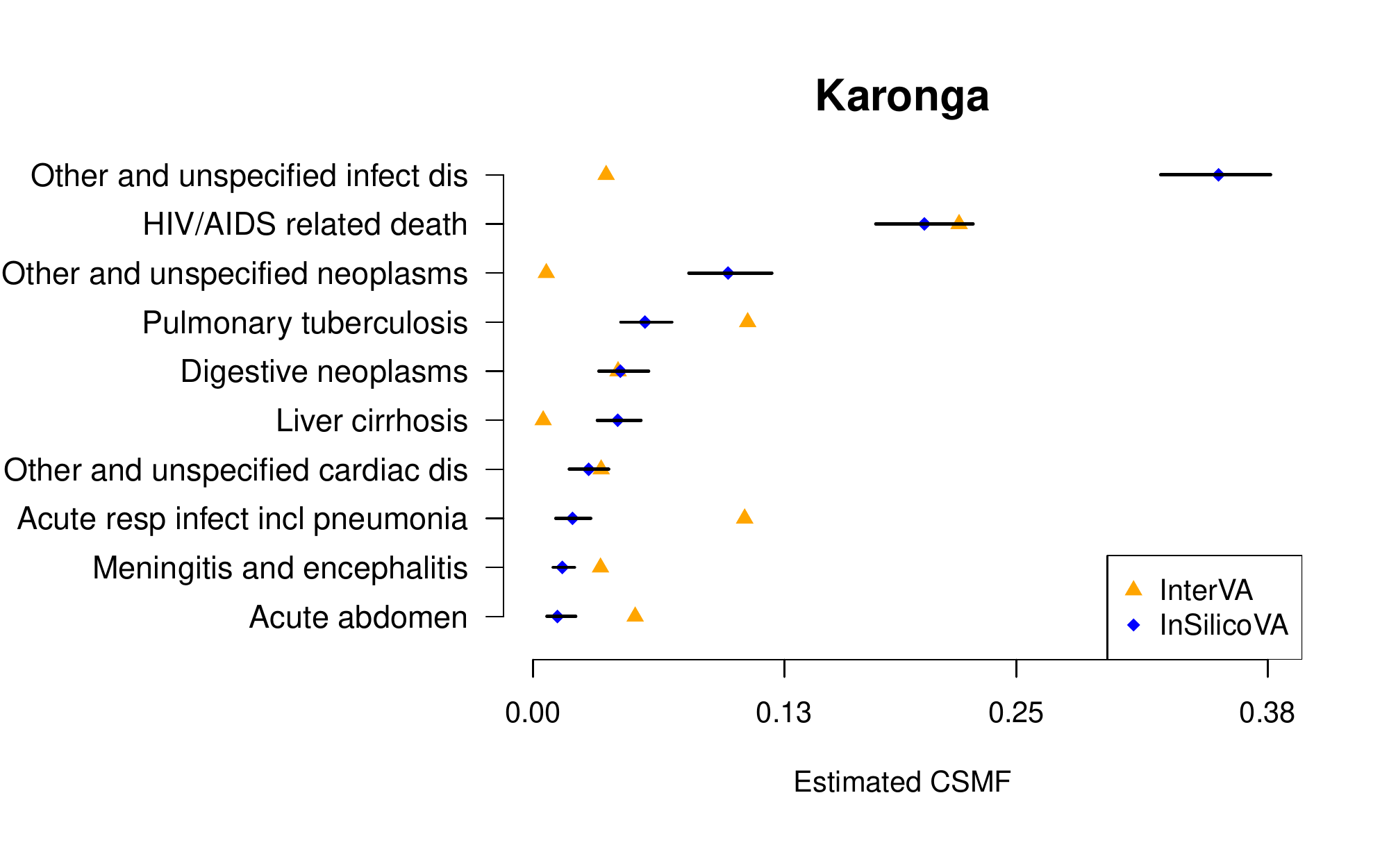}
\captionsetup{format=plain,font=normalsize,margin=0cm,justification=justified}
\caption{{\bf The 10 largest CSMFs.}}
\floatfoot{Estimation comparing InsilicoVA to InterVA in two HDSS sites.  Overall we see that InSilicoVA classifies a larger proportion of deaths into `other/unspecified' categories, reflecting a more conservative procedure that is consistent with the vast uncertainty in these data.  Point estimates represent the posterior mean and intervals are 95\% credible intervals.
}
\label{fig:south}
\end{figure}

The results from  Agincourt and Karonga study are presented in Figure \ref{fig:south}.  InSilicoVA is far more conservative and produces confidence bounds, whereas InterVA does not. A key difference is that InSilicoVA classifies a larger portion of deaths to causes labeled in various ``other'' groups. This indicates that these causes are related to either communicable or non-communicable diseases, but there is not enough information to make a more specific classification.  This feature of InSilicoVA identifies cases that are difficult to classify using available data and may be good candidates for additional attention, such as physician review.  

We view this behavior as a strength of InSilicoVA because it is consistent with the fundamental weakness of the VA approach, namely that both the information obtained from a VA interview and the expert knowledge and/or gold standard used to characterize the relationship between signs/symptoms and causes are inherently weak and incomplete, and consequently it is very difficult or impossible to make highly specific cause assignments using VA.  Given this, we do not want a method that is artificially precise, i.e. forces fine-tuned classification when there is insufficient information.  Hence we view InSilicoVA's behavior as reasonable, ``honest'' (in that it does not over interpret the data) and useful.  ``Useful'' in the sense that it identifies where our information is particularly weak and therefore where we need to apply more effort either to data or to interpretation.


\section{Physician coding}
\label{sec:phys}
The information available across contexts that use VA varies widely.  One common source of external information arises when a team of physicians reviews VA data and assigns a likely cause to each death.  Since this process places additional demands on already scarce physician time, it is only available for some deaths.  Unlike a gold standard dataset where physician information is used as training data for an algorithm, a physician code is typically used as the definitive classification. Physician clarifiers do not have access to the decedent's body to perform a more detailed physical examination, as in a traditional autopsy.   

We propose incorporating physician coded deaths into our hierarchical modeling framework.  This strategy provides a unified inference framework for population cause distributions based on both physician coded and non-physician coded deaths.  We address two challenges in incorporating these additional sources of data.  First, available physician coded data often do not match causes used in existing automated VA tools such as InterVA.  For example in the Karonga HDSS site physicians code deaths in a list of 88 categories and need to be aggregated into broader causes to match InterVA causes.  Second, each physician uses her/his own clinical experience, and often a sense of context-specific disease prevalences, to code deaths, leading to variability and potentially bias.  In Section~\ref{sec:physcode} we present our approach to addressing these two issues.  We then present results in Section~\ref{sec:physresults}.

\subsection{Physician coding framework}
\label{sec:physcode}
In this section we demonstrate how to incorporate physician coding into the InSilicoVA method.  If each death were coded by a single physician using the same possible causes of death as in our statistical approach, the most straightforward means of incorporating this information would be to  replace the $y_{i}$ for a given individual with the physician's code.  This strategy assumes that a physician can perfectly code each death using the information in a VA interview.  In practice it is difficult for physicians to definitively assign a cause using the limited information from a VA interview.   Multiple physicians typically code each death to form a consensus.  Further the possible causes used by physicians do not match the causes used in existing automated methods.  In the data we use, physicians first code based on six broad cause categories, then assign deaths to more specific subcategories.  Since we wish to use clinician codes as an input into our probabilistic model rather than as a final cause determination, we will use the broad categories from the Karonga site.  
Since our data are only for adults we removed the category ``infant deaths.''  
We are also particularly interested in assignment in large disease categories, such as TB or HIV/AIDS, so we add and additional category for these causes.  The resulting list is: (i) non-communicable disease (NCD), (ii) TB or AIDS, (iii) other communicable disease, (iv) maternal cause, (v) external cause or (vi) unknown. The WHO VA standards \citep{who} map the causes of death used in the previous section to ICD-10 codes, which can then be associated with these six broad categories.  We now have a set of broad cause categories $\{1, ..., G\}$ that physicians use and a way to relate these general causes to several causes of death used by InSilicoVA. 

We further assume that we know the probability that a death is related to a cause in each of these broad cause categories. That is, we have a vector of a rough COD distribution for each death: $Z_i = (z_{i1}, ..., z_{iG})$ and $\sum_g z_{ig} = 1$. In situations where each death is examined by several physicians, we can use the distribution of assigned causes across the physicians.  When only one physician reviews the death, we place all the mass on the $Z_{i}$ term representing the broad category assigned by that  one physician. An advantage of our Bayesian approach is that we could also distribute mass across other cause categories if we had additional uncertainty measures. We further add a latent variable $\eta_i \in \{1, ..., G\}$ indicating the category assignment. The posterior of $Y$ then becomes
\[P(y_i | \pi, S_i, Z_i) = \sum_{g = 1}^{G}P(y_i | \boldsymbol{\pi}, \eta_i = g) P(\eta_i = g | Z_i)\]
\noindent Since
$\eta_i = g | Z_i \sim \mbox{Categorical}(z_{i1}, z_{i2}, ..., z_{iG})$ and
$y_i | \boldsymbol{\pi}, \eta_i = g \sim  \mbox{Categorical}(\tilde p_{1i}^{(g)}, \tilde p_{2i}^{(g)},...,\tilde p_{Ci}^{(g)})$
where $p_{ci}^{(g)} \propto p_{1i}^{(NB)} \chi_{cg}$, and $\chi_{cg}$ is the indicator of cause $k$ in category $g$. Without loss of generality we assume each cause belongs to at least one category. Then by collapsing the latent variable $\eta_i$, we directly sample $y_i$ from the posterior distribution:
\[
y_{i}|\pi, S_i, Z_i \sim \mbox{Categorical}(\tilde p_{1i}, \tilde p_{2i}, ...,\tilde p_{Ci})
\]
where \[\tilde p_{ci} = \frac{\sum_{g = 1}^G z_{ig} p_{ci}^{(NB)}}{\sum_{c=1}^C\sum_{g = 1}^G z_{ig} p_{ci}^{(NB)}}. \]

%
%

A certain level of physician bias is inevitable, especially when physicians' training, exposure, and speciality vary. 
Some physicians are more likely to code certain causes than others, particularly where they have clinical experience in the setting and a presumed knowledge of underlying disease prevalences. 
We adopt a two-stage model to incorporate uncertainties in the data and likely bias in the physician codes. First we use a separate model to estimate the bias of the physician codes and obtain the de-biased cause distribution for each death. Then, we feed the distribution of likely cause categories (accounting for potential bias) into InsilicoVA to guide the algorithm. 

For the first stage we used the model for annotation bias in document classification proposed by \citet{salter2013sentiment}.  
The algorithm was proposed to model the annotator bias in rating news sentiment, but if we think of each death as a document with symptoms as words and cause of death as the sentiment categories, the algorithm can be directly applied to VA data with physician coding.  Suppose there are $M$ physicians in total, and for each death $i$ there are $M_i$ physicians coding the cause. Let $Z_i^{(k)} = \left(z_{i1}^{(m)}, z_{i2}^{(m)},..., z_{iG}^{(m)} \right)$ be the code of death $i$ by physician $m$, where $z_{ig}^{(m)} = 1$ if death $i$ is assigned to cause $g$ and 0 otherwise. The reporting bias matrix $\left\{\theta^{(m)}_{gg'}\right\}$ for each physician is then defined as the probability of assigning cause $g'$ when the true cause is $g$.  If we also denote the binary indicator of true cause of death $i$ to be $T_i = \{t_{i1}, t_{i2}, ..., t_{iG}\}$, the conditional probability of observing symptom $j$ given cause $g$ as $p_{j|g}$, and the marginal probability of cause $g$ as $\pi_g$, then the complete data likelihood of physician coded dataset is:
\begin{equation}
\mathcal L( \pi, p , \theta| S, Z, T) = \prod_{i}^{n} \prod_{g}^G \left\{ 
\pi_g \prod_{m}^M \prod_{g'}^G (\theta_{gg'}^{(m)})^{z_{ig'}^{(m)}}  \prod_j^{S} p_{j|g}^{s_{ij}} (1- p_{j|g})^{(1 - s_{ij})}
\right\} ^{T_{ig}}
\label{eqn:phy-lik}
\end{equation} 
We then proceed as in \citet{salter2013sentiment} and learn the most likely set of parameters through an implementation of the EM algorithm. The algorithm proceeds:
\begin{enumerate}
\item For $i=1,...,n$:
\begin{enumerate}
\item initialize $T$ using $\hat t_{ig}  =  \frac{\sum_m^{M_i} z_{ig}^{(m)}}{M_i}$
\item  initialize $\pi$ using $\hat \pi_g = \frac{\sum_{i}t_{ig}}{N}$
\end{enumerate}
\item Repeal until convergence:
\begin{eqnarray*}
\hat \theta_{gg'}^{(m)} &\leftarrow& \frac{\sum_i \hat t_{ig}{z_{ig'}^{(m)}}}{\sum_{g'}\sum_{i}\hat t_{ig}z_{ig'}^{(m)}}\\
\hat p_{jg} &\leftarrow& \frac{\sum_i s_{ij}\hat t_{ig}}{\sum_{i}\hat t_{ig}}\\
\hat \pi_g &\leftarrow& \frac{\sum_i \hat t_{ig}}{N}\\
\hat t_{ig}  &\leftarrow&  
\pi_g \prod_{m}^M \prod_{g'}^G (\theta_{gg'}^{(m)})^{z_{ig'}^{(m)}}  \prod_j^{S} p_{jg}^{s_{ij}} (1- p_{jg})^{(1 - s_{ij})}\\
\hat t_{ig}  &\leftarrow& \frac{\hat t_{ig} }{\sum_{g'}\hat t_{ig'} }. 
\end{eqnarray*}
\end{enumerate}


%


\noindent After convergence, the estimator $\hat t_{ig}$ can then be used in place of $z_{ig}$ in the main algorithm as discussed in Section~\ref{sec:mainmodel}.  
An alternative would be to develop a fully Bayesian strategy to address bias in physician coding.  We have chosen not to do this because the VA data we have is usually interpreted by a small number of physicians who assign causes to a large number of deaths.  Consequently there is a large amount of information about the specific tendencies of each physician, and thus the physician-specific bias matrix can be estimated with limited uncertainty.  A fully Bayesian approach would involve estimating many additional parameters, but sharing information would be of limited value because there are many cases available to estimate the physician-specific matrix.  We believe the uncertainty in estimating the physician-specific matrix is very small compared to other sources of uncertainty.  

\subsection{Comparing results using physician coding}
\label{sec:physresults}
We turn now to results that incorporate physician coding.  We implemented the physician coding algorithm described above on the Karonga dataset described in Section~\ref{sec:realStudy}.  The Karonga site has used physician and clinical officer coding from 2002 to the present. The 1,469 deaths in our data have each been reviewed by physician. Typically each death is reviewed by two physicians; deaths where the two disagree are reviewed by a third.  Over the period of this analysis, 18 physicians reviewed an average of 217 deaths each.
\begin{figure}[htb]
\centering
\includegraphics[width = .7\textwidth]{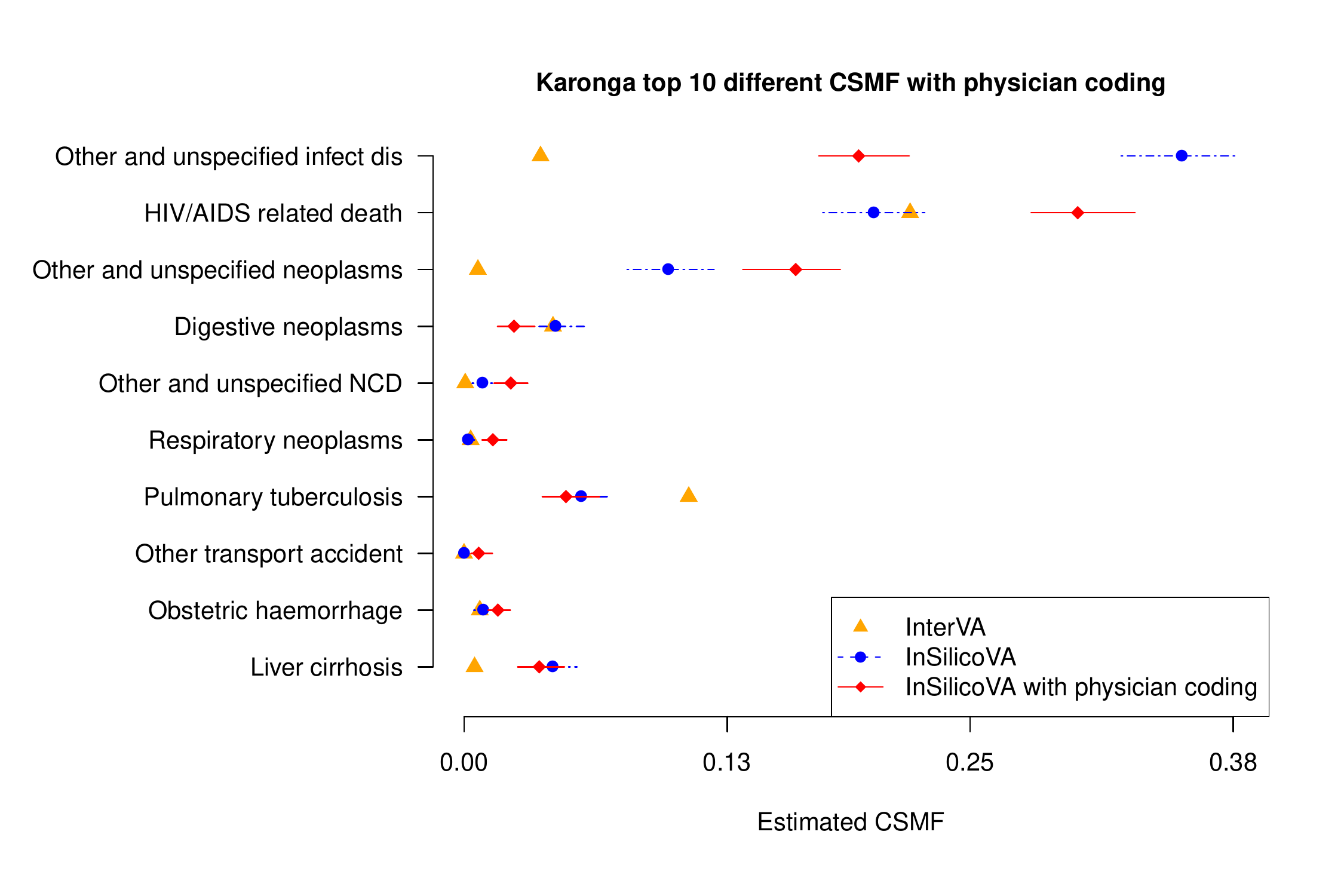}
\captionsetup{format=plain,font=normalsize,margin=2cm,justification=justified}
\caption{{\bf The top 10 most different CSMFs.}}  
\floatfoot{Estimation comparing InterVA with and without physician coding for the Karonga dataset.  InSilicoVA without physician coding categorizes many more deaths as `Other and unspecified infectious diseases' compared to InterVA.  Including physician coding reduces the fraction of deaths in this category, indicating an increase in certainty about some deaths.  Point estimates represent the posterior mean and intervals are 95\% credible intervals.}
\label{fig:phy}
\end{figure}

\begin{figure}[h!]
\centering
\includegraphics[width = .86\textwidth]{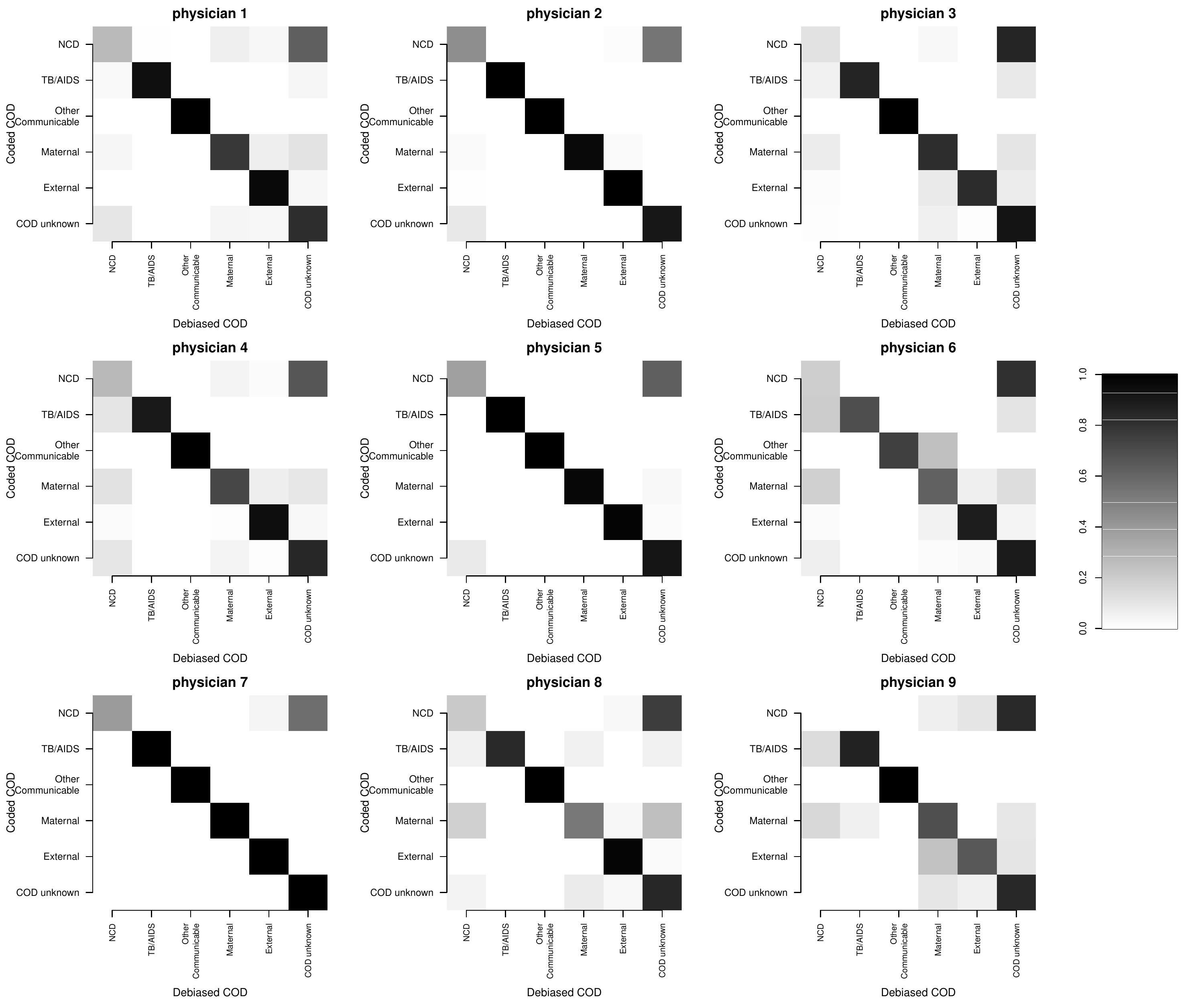}
\captionsetup{format=plain,font=normalsize,margin=1cm,justification=justified}
\caption{\bf{Physician variability.}}  
\floatfoot{Each $6\times6$ square matrix represents a single physician coding verbal autopsy deaths from the Karonga HDSS.  Within each matrix, the shading of each cell corresponds to the propensity of the physician to classify the death into the cause category associated with the cell's row when the the true cause category is the one associated with the cell's column.  The physician bias estimates come from comparing cause assignments for the same death produced by multiple physicians.  A physician with no individual bias would have solid black on the diagonal.  The figure indicates that the variation in both the nature and magnitude of individual physician bias varies substantially between physicians.}
\label{fig:bias}
\end{figure}

We first evaluate the difference in CSMFs estimated with and without incorporating physician-coded causes.  
We use the six broad categories of physician coding described in Section ~\ref{sec:physcode}.
Figure~\ref{fig:phy} compares the CSMFs using InSilicoVA both with and without physician coding.  Including physician coding reduces the fraction of deaths coded as other/unspecified infectious diseases and increases the fraction of deaths assigned to HIV/AIDS.  This is likely the result of physicians with local knowledge being more  aware of the \textit{complete} symptom profile typically associated with HIV/AIDS in their area.  They may also gather useful data from the VA narrative that aids them in making a decision on cause of death.  Having seen multiple cases of HIV/AIDS, physicians can leverage information from combinations of symptoms that is much harder to build into a computational algorithm.  Physicians can also use knowledge of the prevalence of a given condition in a local context.  If the majority of deaths they see are related to HIV/AIDS, they may be more likely to assign HIV/AIDS as a cause even in patients with a more general symptom profile.  

Figure~\ref{fig:bias} shows the estimated physician-specific bias matrices for the nine physicians coding the most deaths.  Since each physician has unique training and experience, we expect that there will be differences between physicians in the propensity to assign a particular cause, even among physicians working in the same clinical context.  Figure~\ref{fig:bias} displays the $\left\{\theta^{(m)}_{gg'}\right\}$ matrix described in Section~\ref{sec:physcode}.  The shading of each cell in the matrix represents the propensity of a given physician to code a death as the cause category associated with its row, given that the true cause category is the cause category associated with its column.  
 If all physicians coded with no individual variation, all of the blocks would be solid black along the diagonal.  Figure~\ref{fig:bias} shows that there is substantial variability between the physicians in terms of the degree of their individual proclivity to code specific causes.  This variation is especially persistent in terms of non-communicable diseases, indicating that physicians' unique experiences were most influential in assigning non-communicable diseases.  
\section{Discussion}
\label{sec:concl}
Assigning a cause(s) to a particular death can be challenging under the best circumstances.  Inferring a cause(s) given the limited data available from sources like VA in many developing nations is an extremely difficult task. In this paper we propose a probabilistic statistical framework for using VA data to infer an individual's cause of death and the population cause of death distribution and quantifying uncertainty in both.  The proposed method uses a data augmentation approach to reconcile individuals' causes of death with the population cause of death distribution.  
We demonstrate how our new framework can incorporate multiple types of outside information, in particular physician codes.  However, many open issues remain.  In our data, we observe all deaths in the HDSS site.  Inferring cause of death distributions at a national or regional level, however, would require an adjustment for sampling.  If sampling weights were known, we could incorporate them into our modeling framework.  In many developing nations, however, there is limited information available to construct such weights.  

\rev{
In practice, the proposed method is intended for situations where ``gold standard'' death are not available and where priors have been elicited from medical experts in the form of the matrix of conditional probabilities ${\bf P}_{s|c}$.  The method is also suitable for situations where there is partial physician coding, especially in situations where the same physician codes a large number of VA surveys.  Through our evaluation experiments using the PHMRC gold standard dataset, we also now suggest that the ${\bf P}_{s|c}$ matrix should be updated to reflect regional and temporal changes in symptom-cause associations.
}

We conclude by highlighting two additional open questions.  For both we stress the importance of pairing statistical models with data collection.  
First, questions remain about the importance of correlation between symptoms in inferring a given cause.  In both InSilicoVA and InterVA the product of the marginal probability of each cause is used to approximate the joint distribution of the entire vector of a decedent's symptoms.  This assumption ignores potentially very informative information about comorbidity between signs/symptoms, i.e. dependence in the manifestation of signs/symptoms.  Physician diagnosis often relies on recognition of a constellation of signs and symptoms combined with absence of others.  Advances in statistical tools for modeling large covariance matrices through factorizations or projections could provide ways to model these interactions.  Information describing these dependencies needs to be present in $\mathbf{P}_{s|c}$, the matrix of conditional probabilities associating signs/symptoms and causes elicited from physicians.  Until now physicians have not been asked to produce this type of information and there does not exist an appropriate data collection mechanism to do this.   When producing the $\mathbf{P}_{s|c}$ matrix, medical experts are only asked to provide information about one cause at a time.  Obtaining information about potentially informative co-occurences of signs/symptoms is essential but will involve non-trivial changes to future data collection efforts.  It is practically impossible to ask physicians about every possible combination of symptoms.  A key challenge, therefore, will be identifying which combinations of symptoms could be useful and incorporating this incomplete association information into our statistical framework.  The $\mathbf{P}_{s|c}$ matrix entries are also currently solicited by consensus and without uncertainty.  Adding uncertainty is straightforward in our Bayesian framework and would produce more realistic estimates of uncertainty for the resulting cause assignment and population proportion estimates.

Second, the current VA questionnaire is quite extensive and requires a great deal of time, concentration and patience to administer.  This burden is exacerbated since interviewees have recently experienced the death of a loved one or close friend.  Further, many symptoms occur either very infrequently or extremely frequently across multiple causes of death.  Reducing the number of items on the VA questionnaire would ease the burden on respondents and interviewers.  This change would likely improve the overall quality of the data, allowing individuals to focus more on the most influential symptoms without spending time on questions that are less informative.  Reducing the number of items on the questionnaire and prioritizing the remaining items would accomplish this goal.  Together the increasing availability of affordable mobile survey technologies and advances in item-response theory, related areas in statistics, machine learning, and psychometrics provide an opportunity to create a more parsimonious questionnaire that dynamically presents a personalized series of questions to each respondent based on their responses to previous questions.


\bibliographystyle{abbrevnamed}
\bibliography{VAbib}

\begin{thebibliography}{}

\bibitem[\protect\citeauthoryear{AbouZahr \bgroup \em et al.\egroup
  }{2007}]{AbouZahr2007}
C.~AbouZahr, J.~Cleland, F.~Coullare, S.~B. Macfarlane, F.~C. Notzon, P.~Setel,
  S.~Szreter, R.~N. Anderson, A.~a. Bawah, A.~P. Betr\'{a}n, F.~Binka,
  K.~Bundhamcharoen, R.~Castro, T.~Evans, X.~C. Figueroa, C.~K. George,
  L.~Gollogly, R.~Gonzalez, D.~R. Grzebien, K.~Hill, Z.~Huang, T.~H. Hull,
  M.~Inoue, R.~Jakob, P.~Jha, Y.~Jiang, R.~Laurenti, X.~Li, D.~Lievesley, A.~D.
  Lopez, D.~M. Fat, M.~Merialdi, L.~Mikkelsen, J.~K. Nien, C.~Rao, K.~Rao,
  O.~Sankoh, K.~Shibuya, N.~Soleman, S.~Stout, V.~Tangcharoensathien, P.~J.
  van~der Maas, F.~Wu, G.~Yang, and S.~Zhang.
\newblock {The way forward.}
\newblock {\em Lancet}, 370(9601):1791--9, November 2007.

\bibitem[\protect\citeauthoryear{Boerma and Stansfi}{2007}]{Boerma}
J.~T. Boerma and S.~K. Stansfi.
\newblock {Health Statistics 1 Health statistics now : are we making the right
  investments ?}
\newblock {\em Tuberculosis}, pages 779--786, 2007.

\bibitem[\protect\citeauthoryear{Byass \bgroup \em et al.\egroup
  }{2012a}]{Byass:2012}
P.~Byass, D.~Chandramohan, S.~Clark, L.~D'Ambruoso, E.~Fottrell, W.~Graham,
  A.~Herbst, A.~Hodgson, S.~Hounton, K.~Kahn, A.~Krishnan, J.~Leitao,
  F.~Odhiambo, O.~Sankoh, and S.~Tollman.
\newblock Strengthening standardised interpretation of verbal autopsy data: the
  new interva-4 tool.
\newblock {\em Global Health Action}, 5(0), 2012.

\bibitem[\protect\citeauthoryear{Byass \bgroup \em et al.\egroup
  }{2012b}]{byass2012strengthening}
P.~Byass, D.~Chandramohan, S.~J. Clark, L.~D'Ambruoso, E.~Fottrell, W.~J.
  Graham, A.~J. Herbst, A.~Hodgson, S.~Hounton, K.~Kahn, et~al.
\newblock Strengthening standardised interpretation of verbal autopsy data: the
  new interva-4 tool.
\newblock {\em Global health action}, 5, 2012.

\bibitem[\protect\citeauthoryear{Byass}{2012}]{byass2012persComm}
P.~Byass.
\newblock Personal communication, 2012.

\bibitem[\protect\citeauthoryear{Byass}{2013}]{2013interVA}
P.~Byass.
\newblock Interva software.
\newblock {\em www.interva.org}, 2013.

\bibitem[\protect\citeauthoryear{Crampin \bgroup \em et al.\egroup
  }{2012}]{crampin2012profile}
A.~C. Crampin, A.~Dube, S.~Mboma, A.~Price, M.~Chihana, A.~Jahn, A.~Baschieri,
  A.~Molesworth, E.~Mwaiyeghele, K.~Branson, et~al.
\newblock Profile: the {K}aronga health and demographic surveillance system.
\newblock {\em International {J}ournal of {E}pidemiology}, 41(3):676--685,
  2012.

\bibitem[\protect\citeauthoryear{Flaxman \bgroup \em et al.\egroup
  }{2011}]{flax}
A.~D. Flaxman, A.~Vahdatpour, S.~Green, S.~L. James, C.~J. Murray, and
  {Consortium Population Health Metrics Research}.
\newblock Random forests for verbal autopsy analysis: multisite validation
  study using clinical diagnostic gold standards.
\newblock {\em Popul Health Metr}, 9(29), 2011.

\bibitem[\protect\citeauthoryear{Gelman and Rubin}{1992}]{gelman1992inference}
A.~Gelman and D.~B. Rubin.
\newblock Inference from iterative simulation using multiple sequences.
\newblock {\em Statistical science}, pages 457--472, 1992.

\bibitem[\protect\citeauthoryear{Gelman \bgroup \em et al.\egroup
  }{1996}]{gelman1996physiological}
A.~Gelman, F.~Bois, and J.~Jiang.
\newblock Physiological pharmacokinetic analysis using population modeling and
  informative prior distributions.
\newblock {\em Journal of the American Statistical Association},
  91(436):1400--1412, 1996.

\bibitem[\protect\citeauthoryear{Hill \bgroup \em et al.\egroup
  }{2007}]{Hill2007b}
K.~Hill, A.~D. Lopez, K.~Shibuya, P.~Jha, C.~AbouZahr, R.~N. Anderson, A.~a.
  Bawah, A.~P. Betr\'{a}n, F.~Binka, K.~Bundhamcharoen, R.~Castro, J.~Cleland,
  F.~Coullare, T.~Evans, X.~{Carrasco Figueroa}, C.~K. George, L.~Gollogly,
  R.~Gonzalez, D.~R. Grzebien, Z.~Huang, T.~H. Hull, M.~Inoue, R.~Jakob,
  Y.~Jiang, R.~Laurenti, X.~Li, D.~Lievesley, D.~M. Fat, S.~Macfarlane,
  P.~Mahapatra, M.~Merialdi, L.~Mikkelsen, J.~K. Nien, F.~C. Notzon, C.~Rao,
  K.~Rao, O.~Sankoh, P.~W. Setel, N.~Soleman, S.~Stout, S.~Szreter,
  V.~Tangcharoensathien, P.~J. van~der Maas, F.~Wu, G.~Yang, S.~Zhang, and
  M.~Zhou.
\newblock {Interim measures for meeting needs for health sector data: births,
  deaths, and causes of death.}
\newblock {\em Lancet}, 370(9600):1726--35, November 2007.

\bibitem[\protect\citeauthoryear{Horton}{2007}]{Horton2007}
R.~Horton.
\newblock {Counting for health.}
\newblock {\em Lancet}, 370(9598):1526, November 2007.

\bibitem[\protect\citeauthoryear{James \bgroup \em et al.\egroup
  }{2011}]{james}
S.~L. James, A.~D. Flaxman, C.~J. Murray, and {Consortium Population Health
  Metrics Research}.
\newblock Performance of the tariff method: validation of a simple additive
  algorithm for analysis of verbal autopsies.
\newblock {\em Popul Health Metr}, 9(31), 2011.

\bibitem[\protect\citeauthoryear{Kahn \bgroup \em et al.\egroup
  }{2012}]{kahn2012profile}
K.~Kahn, M.~A. Collinson, F.~X. G{\'o}mez-Oliv{\'e}, O.~Mokoena, R.~Twine,
  P.~Mee, S.~A. Afolabi, B.~D. Clark, C.~W. Kabudula, A.~Khosa, et~al.
\newblock Profile: Agincourt health and socio-demographic surveillance system.
\newblock {\em International {J}ournal of {E}pidemiology}, 41(4):988--1001,
  2012.

\bibitem[\protect\citeauthoryear{King and Lu}{2008}]{king2}
G.~King and Y.~Lu.
\newblock Verbal autopsy methods with multiple causes of death.
\newblock {\em Statistical Science}, 100(469), 2008.

\bibitem[\protect\citeauthoryear{King \bgroup \em et al.\egroup }{2010}]{king1}
G.~King, Y.~Lu, and K.~Shibuya.
\newblock Designing verbal autopsy studies.
\newblock {\em Population Health Metrics}, 8(19), 2010.

\bibitem[\protect\citeauthoryear{Li \bgroup \em et al.\egroup
  }{2014}]{li2014interva4}
Z.~R. Li, T.~H. McCormick, and S.~J. Clark.
\newblock Interva4: An r package to analyze verbal autopsy data.
\newblock 2014.

\bibitem[\protect\citeauthoryear{Mahapatra \bgroup \em et al.\egroup
  }{2007}]{Mahapatra2007}
P.~Mahapatra, K.~Shibuya, A.~D. Lopez, F.~Coullare, F.~C. Notzon, C.~Rao, and
  S.~Szreter.
\newblock {Civil registration systems and vital statistics: successes and
  missed opportunities}.
\newblock {\em The Lancet}, 370(9599):1653--1663, November 2007.

\bibitem[\protect\citeauthoryear{Maher \bgroup \em et al.\egroup
  }{2010}]{maher2010translating}
D.~Maher, S.~Biraro, V.~Hosegood, R.~Isingo, T.~Lutalo, P.~Mushati, B.~Ngwira,
  M.~Nyirenda, J.~Todd, and B.~Zaba.
\newblock Translating global health research aims into action: the example of
  the alpha network.
\newblock {\em Tropical Medicine \& International Health}, 15(3):321--328,
  2010.

\bibitem[\protect\citeauthoryear{Murray \bgroup \em et al.\egroup
  }{2011a}]{murray}
C.~J. Murray, S.~L. James, J.~K. Birnbaum, M.~K. Freeman, R.~Lozano, A.~D.
  Lopez, and {Consortium Population Health Metrics Research}.
\newblock Simplified symptom pattern method for verbal autopsy analysis:
  multisite validation study using clinical diagnostic gold standards.
\newblock {\em Popul Health Metr}, 9(30), 2011.

\bibitem[\protect\citeauthoryear{Murray \bgroup \em et al.\egroup
  }{2011b}]{murray2011population}
C.~J. Murray, A.~D. Lopez, R.~Black, R.~Ahuja, S.~M. Ali, A.~Baqui, L.~Dandona,
  E.~Dantzer, V.~Das, U.~Dhingra, et~al.
\newblock Population health metrics research consortium gold standard verbal
  autopsy validation study: design, implementation, and development of analysis
  datasets.
\newblock {\em Population health metrics}, 9(1):27, 2011.

\bibitem[\protect\citeauthoryear{Murray \bgroup \em et al.\egroup
  }{2014}]{murray2014using}
C.~J. Murray, R.~Lozano, A.~D. Flaxman, P.~Serina, D.~Phillips, A.~Stewart,
  S.~L. James, A.~Vahdatpour, C.~Atkinson, M.~K. Freeman, et~al.
\newblock Using verbal autopsy to measure causes of death: the comparative
  performance of existing methods.
\newblock {\em BMC medicine}, 12(1):5, 2014.

\bibitem[\protect\citeauthoryear{{R Core Team}}{2014}]{rcitation}
{R Core Team}.
\newblock {\em R: A Language and Environment for Statistical Computing}.
\newblock R Foundation for Statistical Computing, Vienna, Austria, 2014.

\bibitem[\protect\citeauthoryear{Salter-Townshend and
  Murphy}{2013}]{salter2013sentiment}
M.~Salter-Townshend and T.~B. Murphy.
\newblock Sentiment analysis of online media.
\newblock In {\em Algorithms from and for Nature and Life}, pages 137--145.
  Springer, 2013.

\bibitem[\protect\citeauthoryear{Sankoh and Byass}{2012}]{Sankoh2012}
O.~Sankoh and P.~Byass.
\newblock The indepth network: filling vital gaps in global epidemiology.
\newblock {\em International Journal of Epidemiology}, 41(3):579--588, 2012.

\bibitem[\protect\citeauthoryear{Setel \bgroup \em et al.\egroup
  }{2007}]{Setel2007}
P.~W. Setel, S.~B. Macfarlane, S.~Szreter, L.~Mikkelsen, P.~Jha, S.~Stout, and
  C.~AbouZahr.
\newblock {A scandal of invisibility: making everyone count by counting
  everyone.}
\newblock {\em Lancet}, 370(9598):1569--77, November 2007.

\bibitem[\protect\citeauthoryear{Taylor \bgroup \em et al.\egroup
  }{2007}]{twl07}
J.~M.~G. Taylor, L.~Wang, and Z.~Li.
\newblock Analysis on binary responses with ordered covariates and missing
  data.
\newblock {\em Statistics in Medicine}, 26(18):3443--3458, 2007.

\bibitem[\protect\citeauthoryear{Urbanek}{2009}]{rJava}
S.~Urbanek.
\newblock {\em \emph{rJava}: Low-Level {R} to {Java} Interface}, 2009.
\newblock {R}~package version~0.8-1.

\bibitem[\protect\citeauthoryear{{World Health Organization}}{2012}]{who}
{World Health Organization}.
\newblock Verbal autopsy standards: ascertaining and attributing causes of
  death.
\newblock http://www.who.int/healthinfo/statistics/verbalautopsystandards/en/,
  2012.
\newblock Online; accessed 2014-09-08.

\end{thebibliography}


\begin{thebibliography}{}

\bibitem[\protect\citeauthoryear{Byass \bgroup \em et al.\egroup
  }{2012}]{Byass:2012}
P.~Byass, D.~Chandramohan, S.~Clark, L.~D'Ambruoso, E.~Fottrell, W.~Graham,
  A.~Herbst, A.~Hodgson, S.~Hounton, K.~Kahn, A.~Krishnan, J.~Leitao,
  F.~Odhiambo, O.~Sankoh, and S.~Tollman.
\newblock Strengthening standardised interpretation of verbal autopsy data: the
  new interva-4 tool.
\newblock {\em Global Health Action}, 5(0), 2012.

\bibitem[\protect\citeauthoryear{Desai \bgroup \em et al.\egroup
  }{2014}]{desai2014performance}
N.~Desai, L.~Aleksandrowicz, P.~Miasnikof, Y.~Lu, J.~Leitao, P.~Byass,
  S.~Tollman, P.~Mee, D.~Alam, S.~K. Rathi, et~al.
\newblock Performance of four computer-coded verbal autopsy methods for cause
  of death assignment compared with physician coding on 24,000 deaths in
  low-and middle-income countries.
\newblock {\em BMC medicine}, 12(1):20, 2014.

\bibitem[\protect\citeauthoryear{James \bgroup \em et al.\egroup
  }{2011}]{james}
S.~L. James, A.~D. Flaxman, C.~J. Murray, and {Consortium Population Health
  Metrics Research}.
\newblock Performance of the tariff method: validation of a simple additive
  algorithm for analysis of verbal autopsies.
\newblock {\em Popul Health Metr}, 9(31), 2011.

\bibitem[\protect\citeauthoryear{King and Lu}{2008}]{king2}
G.~King and Y.~Lu.
\newblock Verbal autopsy methods with multiple causes of death.
\newblock {\em Statistical Science}, 100(469), 2008.

\bibitem[\protect\citeauthoryear{Maher \bgroup \em et al.\egroup
  }{2010}]{maher2010translating}
D.~Maher, S.~Biraro, V.~Hosegood, R.~Isingo, T.~Lutalo, P.~Mushati, B.~Ngwira,
  M.~Nyirenda, J.~Todd, and B.~Zaba.
\newblock Translating global health research aims into action: the example of
  the alpha network.
\newblock {\em Tropical Medicine \& International Health}, 15(3):321--328,
  2010.

\bibitem[\protect\citeauthoryear{Murray \bgroup \em et al.\egroup
  }{2007}]{murray2007validation}
C.~J. Murray, A.~D. Lopez, D.~M. Feehan, S.~T. Peter, and G.~Yang.
\newblock Validation of the symptom pattern method for analyzing verbal autopsy
  data.
\newblock {\em PLoS Medicine}, 4(11):e327, 2007.

\bibitem[\protect\citeauthoryear{Murray \bgroup \em et al.\egroup
  }{2011a}]{murray}
C.~J. Murray, S.~L. James, J.~K. Birnbaum, M.~K. Freeman, R.~Lozano, A.~D.
  Lopez, and {Consortium Population Health Metrics Research}.
\newblock Simplified symptom pattern method for verbal autopsy analysis:
  multisite validation study using clinical diagnostic gold standards.
\newblock {\em Popul Health Metr}, 9(30), 2011.

\bibitem[\protect\citeauthoryear{Murray \bgroup \em et al.\egroup
  }{2011b}]{murray2011population}
C.~J. Murray, A.~D. Lopez, R.~Black, R.~Ahuja, S.~M. Ali, A.~Baqui, L.~Dandona,
  E.~Dantzer, V.~Das, U.~Dhingra, et~al.
\newblock Population health metrics research consortium gold standard verbal
  autopsy validation study: design, implementation, and development of analysis
  datasets.
\newblock {\em Population health metrics}, 9(1):27, 2011.

\bibitem[\protect\citeauthoryear{Murray \bgroup \em et al.\egroup
  }{2011c}]{murray2011robust}
C.~J. Murray, R.~Lozano, A.~D. Flaxman, A.~Vahdatpour, and A.~D. Lopez.
\newblock Robust metrics for assessing the performance of different verbal
  autopsy cause assignment methods in validation studies.
\newblock {\em Popul Health Metr}, 9(1):28, 2011.

\bibitem[\protect\citeauthoryear{Murray \bgroup \em et al.\egroup
  }{2014}]{murray2014using}
C.~J. Murray, R.~Lozano, A.~D. Flaxman, P.~Serina, D.~Phillips, A.~Stewart,
  S.~L. James, A.~Vahdatpour, C.~Atkinson, M.~K. Freeman, et~al.
\newblock Using verbal autopsy to measure causes of death: the comparative
  performance of existing methods.
\newblock {\em BMC medicine}, 12(1):5, 2014.

\end{thebibliography}

\end{document}